\documentclass[conference, a4paper,10pt]{IEEEtran}
\usepackage[T1]{fontenc}
\pdfoutput=1
\IEEEoverridecommandlockouts

\usepackage[font=footnotesize,caption=false]{subfig} 
\usepackage{tikz}
\usepackage{cite}
\usepackage{tikz}
\usepackage{comment}
\usepackage{cite}
\usepackage{amsmath,amssymb,amsfonts}
\usepackage{soul}
\usepackage{algorithmic}
\usepackage{enumitem}
\usepackage{tikz}
\allowdisplaybreaks
\usepackage{pgfplots}
\pgfplotsset{width=14cm,compat=1.9}
\usepackage[noend,ruled,algo2e]{algorithm2e}
\usepackage{geometry}
\usetikzlibrary{arrows}
\usetikzlibrary{arrows.meta}
\usetikzlibrary{automata, positioning}
\usepackage{algorithm, algorithmic}
\usepackage{graphicx}
\usepackage{textcomp}

\usepackage{xcolor}
\usepackage{booktabs}
\usepackage[short]{optidef}
\usepackage{mathtools}
\usepackage{glossaries}
\usepackage{subfig}
\usepackage{algorithm}
\usepackage{bbm}
\usepackage{multirow}

\usepackage{mathtools}
\usepackage{mathtools,amsthm}
\usepackage{bm}
\newtheorem{theorem}{Theorem} 
\newtheorem{remark}{Remark}

\newtheorem{lemma}{Lemma}
\newtheorem{Proposition}{Proposition}

\newtheorem{corollary}{Corollary}

\def\BibTeX{{\rm B\kern-.05em{\sc i\kern-.025em b}\kern-.08em
    T\kern-.1667em\lower.7ex\hbox{E}\kern-.125emX}}

\makeatletter
\def\blfootnote{\xdef\@thefnmark{}\@footnotetext}
\makeatother

\definecolor{color3}{HTML}{FFD700}
\definecolor{color2}{HTML}{EA5F94}
\definecolor{color1}{HTML}{9D02D7}
\definecolor{color0}{HTML}{0000FF}

\newcommand{\remove}[1]{}

\usepackage{xcolor,cancel}

\title{On the Regret of Coded Caching \\ with Adversarial Requests}
\remove{\author{Anonymous authors}}
\author{\IEEEauthorblockN{Anupam Nayak}
\IEEEauthorblockA{IIT Bombay \\
anupam@ee.iitb.ac.in}
\and
\IEEEauthorblockN{Kota Srinivas Reddy}
\IEEEauthorblockA{IIT Madras\\
ksreddy@ee.iitm.ac.in}  
\and
\IEEEauthorblockN{Nikhil Karamchandani}
\IEEEauthorblockA{IIT Bombay \\
nikhilk@ee.iitb.ac.in}  
}

\begin{document}
\maketitle

\begin{abstract}
We study the well-known coded caching problem in an online learning framework, wherein requests arrive sequentially, and an online policy can update the cache contents based on the history of requests seen thus far. We introduce a caching policy based on the Follow-The-Perturbed-Leader principle and show that for any time horizon $T$ and any request sequence, it achieves a sub-linear regret of $\mathcal{O}(\sqrt{T})$ with respect to an oracle that knows the request sequence beforehand. Our study marks the first examination of adversarial regret in the coded caching setup. Furthermore, we also address the issue of switching cost by establishing an upper bound on the expected number of cache updates made by our algorithm under unrestricted switching and also provide an upper bound on the regret under restricted switching when cache updates can only happen in a pre-specified subset of timeslots. Finally, we validate our theoretical insights with numerical results using a real-world dataset.
\remove{
within the framework of online learning theory. Specifically, our focus is on minimizing the regret of a coded caching problem with adversarial requests. To our knowledge, our study marks the first examination of adversarial regret in coded caching setup with a broadcast channel and coded delivery. To minimize the regret, we introduce a policy based on Follow-The-Perturbed-Leader. Through comparative analysis between our algorithm and a static oracle, we demonstrate that our policy achieves sub-linear regret of $\mathcal{O}(\sqrt{T})$. Furthermore, we also address the issue of switching costs by establishing an upper bound on the expected number of switches in our algorithm under unrestricted switching conditions, and also provide an upper bound on the regret under restricted switching. Additionally, we validate our theoretical insights with numerical results on a real-world Movielens 1M dataset.
}
\end{abstract}

\section{Introduction}
\vspace{-4pt}
\blfootnote{N. Karamchandani's work was supported by SERB grants on `Online Learning with Constraints" and `Leveraging Edge Resources for Service Hosting", and  a SERB MATRICS grant. Kota Srinivas Reddy's work was supported by the Department of Science and Technology (DST), Govt. of India, through the INSPIRE faculty fellowship.}
The unprecedented surge in demand for high-definition content over the internet has resulted in an increased load on the underlying communication networks. This challenge can be effectively mitigated through the widespread adoption of Content Delivery Networks (CDNs). CDNs strategically deploy storage devices or caches across large geographical regions. During off-peak hours, these caches are utilized to proactively pre-fetch popular content \cite{maddah2014fundamental}. This proactive strategy aims to reduce network traffic during peak hours when users generate the highest volume of requests.



In the realm of caching, traditional policies emphasizing local caching gains have been extensively investigated in the literature, as exemplified by \cite{wessels2001web} and the associated references. More recently, Maddah-Ali and Niesen \cite{maddah2014decentralized,maddah2014fundamental} delved into cache networks, introducing the concept of `coded caching' which allows for coding of information while delivering content to users.   Their work resulted in policies that not only achieve significant local caching gains but also offer substantial global caching gains. \remove{Furthermore, the policies proposed in \cite{maddah2014decentralized,maddah2014fundamental} are order-optimal with respect to an information-theoretic lower bound.}

The field of coded caching has since become a vibrant area of research within information theory, exploring various facets of cache networks, including network topology \cite{hachem2017coded,karamchandani2016hierarchical,ji2015fundamental}, content popularity \cite{zhang2017coded,deng2022fundamental,hachem2017coded,hachem2015effect,niesen2016coded,sahraei2019optimal,ji2017order}, and security and privacy \cite{gurjarpadhye2022fundamental,yan2021fundamental,ravindrakumar2017private,wan2020coded,zewail2019device}. In general, these works propose content placement and delivery schemes and subsequently evaluate their performance against the information-theoretic lower bound, often demonstrating a gap of at most a constant multiplicative factor, independent of the system size.


In our paper, we focus on coded caching in the framework of online learning theory. As previously mentioned, existing works \cite{zhang2017coded,deng2022fundamental,hachem2017coded,hachem2015effect,niesen2016coded,sahraei2019optimal,ji2017order} address content popularity framework and devise policies by assuming known and static content popularity. In practical scenarios, the actual content popularity may be unknown, necessitating an emphasis on online learning policies for coded caching where the actions of the caching policy at each time are based on the history of actions and observations. To judge the performance of an online caching policy, we focus on the notion of \textit{adversarial regret} \cite{torlat}, which considers the worst-case (over all possible request sequences) additive gap between the performance of the policy and an (static) oracle that knows all the requests beforehand.  

{To the best of our knowledge, online learning policies for coded caching have been previously explored only in \cite{pedarsani2015online,peter2021decentralized,nayak2023regret}. The setting explored in\cite{pedarsani2015online,peter2021decentralized} involved users requesting files from a catalog that evolved (slowly) over time. The aim was to devise a cache update rule while preserving the benefits of coded caching. Alongside differences in the problem formulation, another key difference in our work is the choice of performance metric;  we consider the worst-case regret (additive gap), while \cite{pedarsani2015online,peter2021decentralized} focus on a multiplicative gap. The setting in \cite{nayak2023regret} is closer to our work, wherein the goal is to minimize the regret in a stochastic setting, assuming that content popularity follows a static probability distribution, which is a priori unknown to the learner.} \remove{The objective is to design a policy that dynamically learns popularity and achieves sub-linear regret}

\remove{In contrast, our focus lies in regret minimization within the adversarial setting, where our aim is to design online policies that minimize regret across all possible request patterns. The adversarial setup holds practical significance due to two key factors: (\emph{i}) content popularity may not adhere to any specific distribution, and (\emph{ii}) content popularity may be dynamic rather than static. We also show that the policy designed here achieves a constant regret in the stochastic setting. Additionally, we use the exact expected rate expression, non-linear in policy parameters and observations, unlike \cite{nayak2023regret}, which uses an approximate, linear rate expression. Apart from coded delivery, the primary technical challenge arises when incorporating a broadcast channel with multiple users ($K$). In this scenario, the rate expression becomes more complex, deviating from a straightforward linear counting problem where the total rate is the sum of individual rates across all $K$ channels.}

In contrast, our focus lies in regret minimization within the adversarial setting, where our aim is to design online policies that minimize the maximum regret across all possible request patterns. The adversarial setup holds practical significance as content popularity may not adhere to any specific distribution and may be dynamic rather than static. 
Additionally, we use the exact expected rate expression, non-linear in policy parameters and observations, unlike \cite{nayak2023regret}, which uses an approximate, linear rate expression. Apart from coded delivery, the primary technical challenge arises when incorporating a broadcast channel with multiple users ($K$). In this scenario, the rate expression becomes more complex, deviating from a straightforward linear counting problem where the total rate is the sum of individual rates across all $K$ channels.

The exploration of caching systems within the online learning framework has recently garnered attention, particularly in the context of a single cache. Motivated by recent advancements in online convex optimization \cite{cohen2015following,abernethy2014online,zinkevich2003online}, several online caching policies have been proposed, including Online Gradient Ascent \cite{paschos2019learning,paschos2019learning-2,paschos2020online}, Online Mirror Descent \cite{salem2021no}, and Follow the Perturbed Leader (FTPL) \cite{zarin2022regret,mukhopadhyay2021online,bhattacharjee2020fundamental,paria2021texttt}. Notably, these approaches specifically target adversarial requests, showcasing the achievement of an order-optimal regret of $\mathcal{O}(\sqrt{T})$, where $T$ is the time horizon. In a similar vein, our work takes a preliminary step in investigating the popular coded caching framework within the context of online learning.

\subsection*{Novelty and Contributions}Our work focuses on the adversarial regret version of the coded caching problem. To the best of our knowledge, our work is the first to study adversarial regret in a coded caching setup with a broadcast channel and coded delivery. Similar to \cite{zarin2022regret,mukhopadhyay2021online,bhattacharjee2020fundamental,paria2021texttt}, we employ an FTPL-based algorithm for our coded caching problem and demonstrate its ability to achieve $\mathcal{O}(\sqrt{T})$ regret. A key technical challenge arises from the non-linear nature of the expected rate expression in the coded caching scenario, which complicates the analysis\footnote{The work in \cite{paria2021texttt} considered bipartite caching networks, for which the rate expression was not linear in terms of the cache configuration. They dealt with it by switching to a `virtual' action space where the reward function was indeed linear.}.  We resolve this issue by rewriting the rate expression using a careful transformation of the request vector.
\remove{\begin{itemize}
    \item For an integer $n$, $[n]$ denotes the set $\{1,2,...,n\}$.
    \item For a given collection of scalars/vectors $r_i$ such that $i\in\mathbb{N}$ and an integer $t$, $(r_i)_{i=1}^{t}$ denotes the multi-set $\{r_1,r_2,...,r_t\}$.
    \item For $N-$dimensional vectors $\mathbf{a}=[a_1,a_2,...,a_N]^T$ and $\mathbf{b}=[b_1,b_2,...,b_N]^T$, $\langle \mathbf{a}, \mathbf{b}\rangle$ denotes the dot product between the vectors $\mathbf{a}$ and $\mathbf{b}$. 
\end{itemize}}



\section{Problem Formulation}\label{sec:sysmodel}
\begin{figure}
    \centerline{\scalebox{.59}{\includegraphics{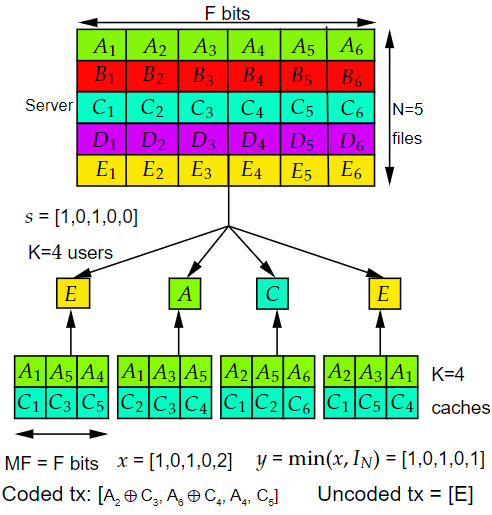}}}
  \caption{\sl An illustration of coded caching problem setup.  It contains a central server with $N=5$ files, each of size $F$ bits, and $K=4$ users, connected to a separate cache of size $MF=F$ bits. Here $s=[1,0,1,0,0]$ indicates that \textit{stored files} are $A$ and $C$, and \textit{unstored files} are the remaining. For a request profile $\mathbf{r}=(E,A,C,E)$, it's request pattern is $x=[1,0,1,0,2]$.}
    \vspace{-0.2in}
    \label{Fig:System}
  \end{figure}
  Consider the system shown in Figure \ref{Fig:System}, which contains a remote backend server that holds $N$ files, each of size $F$ bits, and is connected to $K$ users via an error-free broadcast medium. Each user seeks files from the server and is connected to a local cache storage of size $MF$ bits. Typically, the cache size $M$ is much smaller compared to the number of files $N$. Time is divided into slots with horizon $T \in \mathbb{N}$. Each slot contains two phases, namely the placement phase and the delivery phase.

For every $t \in [T]$, in the placement phase, the caches store content related to the $N$ files. Let $z_t^k$ denote the cache content stored at cache/user $k$ at time $t$. The cache content may change over time and will be a function of the file requests observed in the past as well as the history of cached content. After the placement phase, we get a set of file requests, one from each user, which marks the start of the delivery phase. Let $\mathbf{r}_t = (r_t^1, r_t^2, ..., r_t^K)$ denote the file request profile at time $t$, where $r_t^k$ denotes the file requested by User $k$ in slot $t$.
Based on the users' request profile $\mathbf{r}_t$ and the cached content profile $\mathbf{z}_t = (z_t^1, z_t^2, ..., z_t^K)$, the central server transmits a message $\mathcal{X}_t$ of size/rate $K_{\mathbf{r}_t}(t)$ such that every user $k$ can recover its requested file $r_t^k$ using the server's transmitted message $\mathcal{X}_t$ and its cached content $z_t^k$. We aim to design an algorithm $\pi$, which determines the cache placement and content delivery schemes in each time slot, such that the server's cumulative transmission rate given by $K^{\pi}(T) := \sum_{t=1}^T K_{\mathbf{r}_t}(t)$ is minimized.

\remove{
{\color{red}Similar to standard adversarial settings studied in learning theory, we compare the performance of any algorithm $\pi$ with that of a static oracle, which knows the entire request profile from time $t = 1$ to $t = T$ and uses a fixed placement over the entire time horizon $T$, chosen to minimize the cumulative transmission rate. However, the exact oracle problem described above remains unsolved; that is, the optimal cache contents for each user and the optimal content delivery mechanism, having seen the entire requests at each user in a broadcast network, are not known. Nevertheless, a closely related variant, "coded caching under arbitrary popularity distribution" \cite{zhang2017coded}, has been researched. Hence, we consider the results in \cite{zhang2017coded} as a benchmark for the oracle's algorithm.}
}

\remove{The main theme of the policy proposed in \cite{zhang2017coded} is that during the placement phase, files are divided into two categories: \textit{stored files} and \textit{unstored files} depending on the file popularity. Then, the \textit{stored files} are stored according to the decentralized placement policy proposed in \cite{maddah2014decentralized}. During the delivery phase, requests corresponding to the \textit{unstored files} are served using the server's broadcast channel, and requests corresponding to the \textit{stored files} are served according to the delivery method proposed in \cite{maddah2014decentralized}. Zhang et al. showed that the proposed policy in \cite{zhang2017coded} is order-wise optimal with respect to an information-theoretic lower bound. Motivated by this, as a first step to the coded caching problem with adversarial regret, we restrict our analysis to the algorithms that follow placement and delivery policies similar to those in \cite{zhang2017coded}. That is, we consider the algorithms such that in each time slot $t$, based on the request patterns $(\mathbf{r}_i)_{i=1}^{t-1}$, the algorithm chooses a set of files as \textit{stored files} and then stores the \textit{stored files} according to the placement policy in \cite{maddah2014decentralized}. During the delivery phase, these algorithms serve the requests corresponding to the \textit{unstored files} using the server's broadcast channel and serve the requests corresponding to the \textit{stored files} according to the delivery method proposed in \cite{maddah2014decentralized}. Hence, the main task of an algorithm is to choose a set of files as \textit{stored files}, based on the history of request patterns $(\mathbf{r}_i)_{i=1}^{t-1}$. The formal and full description of our algorithms` placement and delivery policies at each time slot is given below:}
As is standard in online learning theory \cite{cohen2015following}, we compare the performance of any online algorithm $\pi$ with that of a static oracle, which knows beforehand the entire request profile from time $t = 1$ to $t = T$ and uses the best static (fixed) placement over the entire time horizon $T$ along with an optimal delivery policy\footnote{For a given user request profile and the content placement, finding the optimal delivery policy is equivalent to solving an \textit{index coding} problem \cite{reddy2021structured}.} to minimize the cumulative transmission rate. However, the characterization of such an oracle, which includes identifying an optimal placement and delivery policy for every collection of requests, is computationally hard and remains an open problem. Therefore, for analytical tractability and as an initial approach to address the adversarial regret in coded caching problem, we concentrate on a restricted class of policies. Besides having a very natural structure, these policies are also akin to the approach proposed in \cite{zhang2017coded}, which is order-wise optimal for the "coded caching under (known) arbitrary popularity distribution" problem with respect to an information-theoretic lower bound.


\remove{
{\color{green}Now, we discuss the proposed policy in \cite{zhang2017coded}. During the placement phase, files are divided into two categories: \textit{stored files} and \textit{unstored files}, depending on the file popularity. Then, the \textit{stored files} are stored according to the decentralized placement policy proposed in \cite{maddah2014decentralized}. During the delivery phase, requests corresponding to the \textit{unstored files} are served using the server's broadcast channel, and requests corresponding to the \textit{stored files} are served according to the delivery method proposed in \cite{maddah2014decentralized}. For a given arbitrary popularity distribution, \cite{zhang2017coded} showed that the proposed policy is order-wise optimal with respect to an information-theoretic lower bound. Motivated by this, as a first step to the coded caching problem with adversarial regret, we restrict our analysis to the algorithms that follow placement and delivery policies similar to those in \cite{zhang2017coded}. That is,} 
}
Next, we first briefly describe this class of policies and then give details of the placement and delivery phases of any such policy later. In each time slot $t$, based on the history of request patterns $(\mathbf{r}_i)_{i=1}^{t-1}$, the scheme divides the set of files into two categories: \textit{stored files} and \textit{unstored files}. The \textit{stored files} are then cached across the users according to the placement policy in \cite{maddah2014decentralized}. During the delivery phase, the requests corresponding to the \textit{unstored files} are directly handled by the server using the broadcast channel while  the requests corresponding to the \textit{stored files} are served according to the delivery method proposed in \cite{maddah2014decentralized}. Note that any scheme in this class is characterized by its choice of \textit{stored files} in each slot. The formal and full description of the schemes' placement and delivery policies at each time slot is given below:


\begin{enumerate}
     \item \textbf{Placement phase:}
Let a binary vector $\mathbb{I}_N = [1,1,\ldots, 1]$ denotes an $N$-dimensional vector with each entry being 1. Let a binary vector $\mathbf{s} \in \{0,1\}^N$ represent a cache configuration, where an entry with a value of 1 represents a \textit{stored file} and an entry with a value of 0 represents an \textit{unstored file}. To avoid underutilization of caches, a cache configuration $\mathbf{s} \in \{0,1\}^N$ is said to be feasible if it contains more than $M$ entries with a value of 1. Let the set $\mathcal{S}$ denotes the collection of all feasible cache configurations. At time $t$, the algorithm selects a subset of files $\mathbf{s}_t \in \mathcal{S}$ as the cache configuration. Then, each user independently and randomly chooses equal fractions ($\frac{MF}{\langle \mathbf{s}_t, \mathbb{I}_N \rangle}$ bits) of all the files with a "1" entry in $\mathbf{s}_t$ to populate their caches. It's worth noting that the placement policy also mandates that the subset $\mathbf{s}_t$ remains the same across all users. However, the users independently sample bits, resulting in different random fractions of the same set of files being stored in their caches.

    \item \textbf{Delivery phase:} After the placement phase, users reveal their corresponding requests and the delivery phase starts. Let $\mathbf{x}_t = [{x}_t(1), {x}_t(2), \ldots, {x}_t(N)]$ be an $N$-dimensional vector where $x_t(i)$ denotes the number of users requesting file $i$ at time slot $t$. Note that $\langle \mathbf{x}_t, \mathbb{I}_N \rangle = K$, the total number of users. Let $\mathbf{y}_t = \min\{\mathbb{I}_N, \mathbf{x}_t\}$, a pointwise minimum of  $\mathbf{x}_t$ with a vector of all ones, denotes whether a file was requested by at least one user. Once the users make the requests, the delivery at time $t$ takes place using the following two steps.\\
    \textbf{Uncoded Transmission: }
    The requests corresponding to the "unstored files" at time $t$ (i.e., the requests for the files with a "0" entry in $\mathbf{s}_t$) are served directly by the server through uncoded transmission. Since we have a broadcast medium, we only need one transmission for these files even if multiple users request them. Thus, the length of the uncoded transmission is given by the inner product of $\mathbf{y}_t$ with the set of files not cached $(\mathbb{I}_N-\mathbf{s}_t) = \langle (\mathbb{I}_N - \mathbf{s}_t), \mathbf{y}_t \rangle$.\footnote{Note that the actual transmission length is $\langle (\mathbb{I}_N - \mathbf{s}_t), \mathbf{y}_t \rangle|F|$. But, we will be dealing with a normalized size.}

\textbf{Coded Transmission:}
The remaining requests (i.e., the requests for the files with a "1" entry in $\mathbf{s}_t$) are served jointly by the cache contents and a coded message from the server. The coded message design follows the decentralized coded caching\cite{maddah2014decentralized} given the subset of the files to be cached is $\mathbf{s}_t$ at time $t$.
 Let $U_1$ denote the set of users that request files from the cached set $\mathbf{s}_t$, $|U_1| = \langle \mathbf{x}_t,\mathbf{s}_t\rangle$.
For every subset $u \in U_1$ with $|u|\neq 0$, transmit $\bigoplus_{k\in u} V_{k,u\setminus\{k\}}$.
Here, $V_{k,u\setminus\{k\}}$ denotes all the bits that are requested by user $k\in u$, are present in the cache of all users in $u\setminus\{k\}$, and that are not stored in the caches of any other user in $U_1\setminus u$. Note that using the above coded message and the content stored in its caches, every user can recover their corresponding requested file, see \cite{zhang2017coded} for more details about the decoding.

Therefore, the expected length of the coded transmission is equal to the expected sum of lengths of all these messages $\bigoplus_{k\in u} V_{k,u\setminus\{k\}}$ for all $u\subseteq U_1$, where the length of a message $\bigoplus_{k\in u} V_{k,u\setminus\{k\}}$ is $|\bigoplus_{k\in u} V_{k,u\setminus\{k\}}| =\max_{k\in u} | V_{k,u\setminus\{k\}}|$.

Note that by using the above coded and uncoded transmissions, every user can recover their requested files, See \cite[Theorem 1 proof]{maddah2014decentralized} for more details. 
\end{enumerate}

The following proposition gives the message size for a given cache configuration $\mathbf{s}_t$ and the request pattern $\mathbf{x}_t$.  
\begin{Proposition}\label{proposition:expected_rate}
    For a coded caching problem with the given cache configuration $\mathbf{s}_t$ and the request vector $\mathbf{x}_t$, the above-discussed placement and delivery policies give a transmission rate of expected length $K(\mathbf{s}_t,\mathbf{x}_t)$ given by
\begin{multline}
        K(\mathbf{s}_t,\mathbf{x}_t) =\underbrace{\langle(\mathbb{I}_N-\mathbf{s}_t),\mathbf{y}_t\rangle}_{\text{Uncoded transmission}} +\\ \underbrace{\left(\frac{\langle \mathbf{s}_t, \mathbb{I}_N\rangle}{M}-1\right)\left(1-\left(1-\frac{M}{\langle \mathbf{s}_t, \mathbb{I}_N\rangle}\right)^{\langle \mathbf{x}_t,\mathbf{s}_t\rangle}\right)}_{\text{Coded transmission}}, 
        \label{eq:ratexp1}
\end{multline}
where $\mathbf{y}_t=\min\{\mathbb{I}_N, \mathbf{x}_t\}$.
\end{Proposition}
\begin{proof}
Derivation of expected coded transmission length provided in Appendix of the full paper \cite{nayak2024ontheregret}.
\end{proof}
 Note that $K(\mathbf{s}_t,\mathbf{x}_t)$, given in \eqref{eq:ratexp1}, depends on the requests only through $\mathbf{x}_t$. So the knowledge of exact request profile $\mathbf{r}_t = (r_t^1, r_t^2, \ldots, r_t^K)$ is not necessary to compute $K(\mathbf{s}_t,\mathbf{x}_t)$. Thus, knowing the request pattern $\mathbf{x}_t$ is sufficient to choose the cache configuration ($\mathbf{s}_t$) for the policy at time $t$. Equation \eqref{eq:ratexp1} can be rewritten as follows:

\begin{equation}
    K(\mathbf{s}_t,\mathbf{x}_t) = \underbrace{\left\langle\left(\mathbf{s}_t-\frac{M}{N}\mathbb{I}_N\right), f(\mathbf{x}_t,\mathbf{s}_t)-\mathbf{y}_t\right\rangle}_{T_0} +  h(\mathbf{x}_t),
    \label{eq:rateexp_modified}
\end{equation}
where $f(\mathbf{x}_t,\mathbf{s}_t) = \frac{1}{M}\left(1-\left(1-\frac{M}{\langle \mathbf{s}_t, \mathbb{I}_N\rangle}\right)^{\langle \mathbf{x}_t,\mathbf{s}_t\rangle}\right)\mathbb{I}_N $ and $h(\mathbf{x}_t) = \left(1- \frac{M}{N}\right)\langle \mathbf{y}_t, \mathbb{I}_N\rangle$. Note that in \eqref{eq:rateexp_modified}, for given a request pattern $\mathbf{x}_t$, $h(\mathbf{x}_t)$ is a constant and $T_0$ is the only part that can be minimized using an appropriate choice of $\mathbf{s}_t$. In order to minimize the cumulative transmission rate, every algorithm chooses a cache configuration $\mathbf{s}_t \in \mathcal{S}$, based on the history of request patterns $(\mathbf{x}_i)_{i=1}^{t-1}$, in each time slot $t$. After that, placement and delivery phases occur according to the above description. For a given request pattern $(\mathbf{x}_t)_{t=1}^T$, if an algorithm $\pi$'s cache configuration is $(\mathbf{s}_t)_{t=1}^T$, then its cumulative expected transmission rate is given by
\begin{align}\label{eqn:algortim_performace_general}
    K^{\pi}(T) = \sum_{t=1}^T \mathbb{E}[K(\mathbf{s}_t,\mathbf{x}_t)],
\end{align}
where the expectation is with respect to the randomness generated by the algorithm and the storage of random bits involved in the placement phase of the policy.

\remove{As mentioned earlier, we compare the performance of any algorithm $\pi$ with that of a static oracle that knows the entire request profile (equivalent to knowing the request pattern $\mathbf{x}_t$ as far as rate minimization is concerned), from time $t = 1$ to $t = T$ and uses a fixed placement over the entire time horizon $T$, chosen to minimize the cumulative transmission size. The aim is to propose an optimal algorithm that works in any adversarial setting and gives sublinear regret. 
For a given request pattern $(\mathbf{x}_t)_{t=1}^T$, if an algorithms $\pi$'s cache configuration is $(\mathbf{s}_t)_{t=1}^T$, then its cumulative expected transmission rate is 
\begin{align}\label{eqn:algortim_performace_general}
    K^{\pi}(T)=\sum_{t=1}^T\mathbb{E}[K(\mathbf{s}_t,\mathbf{x}_t)],
\end{align}
where the expectation is w.r.t. the randomness generated by the algorithm and the storage of random bits involved in the placement phase of the policy.}

To evaluate an algorithm's performance, we compare its cumulative expected transmission rate given in \eqref{eqn:algortim_performace_general} with the performance of a static oracle which knows the entire request pattern profile from time $t = 1$ to $t = T$ and uses a fixed cache configuration over the entire time horizon $T$, chosen to minimize the cumulative transmission size. Therefore, for a given request pattern $(\mathbf{x}_t)_{t=1}^T$, the oracle performance is given by 
\begin{align}\label{eqn:oracle_performace_general}
    K^{\text{o}}(T)=\min_{\mathbf{s}\in \mathcal{S}}\sum_{t=1}^T\mathbb{E}[K(\mathbf{s},\mathbf{x}_t)],
\end{align}
where $\mathcal{S}$ is the collection of all feasible cache configurations. 
Thus, for a request pattern $(\mathbf{x}_t)_{t=1}^T$, the regret of an online algorithm $\pi$ is given by 
\begin{align}\label{eqn:algortim_regret_general} 
  R_{\pi}((\mathbf{x}_t)_{t=1}^T,T):=K^{\pi}(T) - K^{\text{o}}(T).
\end{align} In adversarial regret minimization problems, no assumptions are made about the requests. Therefore, an algorithm's adversarial regret is calculated for the worst case, and is given by
\begin{align}\label{eqn:algortim_regret_adversarial}
  R_{\pi}(T) =  \max_{\substack{\mathbf{x}_t \in X, \forall t}} R_{\pi}((\mathbf{x}_t)_{t=1}^T,T),
\end{align}
where  $X$ represents the set of all feasible $\mathbf{x}_t$ ($\mathbf{x} | x(i)\geq 0, \langle \mathbf{x}, \mathbb{I}_N\rangle = K$), i.e., all feasible request patterns.

Our aim is to design an algorithm $\pi$ that achieves the minimum worst-case regret $R_{\pi}(T)$ with respect to the static oracle. With this aim, in the rest of our paper, we propose an algorithm (Algorithm \ref{alg:FTPL}) and show that our algorithm achieves $\mathcal{O}(\sqrt{T})$ adversarial regret. 

An algorithm that fetches a large number of files into the caches each time to change the cache configuration is not ideal as fetching files into the caches causes latency and consumes bandwidth. Thus, we also consider the switching cost issue. Let $C_{\pi}(T)$ denotes the expected number of cache configuration switches until time $T$ for an Algorithm $\pi$, then $C_{\pi}(T)$ is given by
\begin{equation*}
    C_{\pi}(T) = \mathbb{E}\left[\sum_{t=1}^{T-1}\mathbb{I}(\mathbf{s}_{t+1}\neq \mathbf{s}_t)\right].
\end{equation*}
We address the issue of switching costs using two results. Firstly, we provide an upper bound on the expected number of switches incurred by our algorithm for the unrestricted switching case. We also provide an upper bound on the regret incurred by our proposed algorithm when the cache configuration is allowed to switch only in a set of predefined arbitrary $L$ restricted slots given by $\mathcal{T}=\{t_i:i\in[L], t_0=0, t_L=T; t_i\in \mathbb{N}, 0\leq t_{i-1}<t_i\leq T\}$. Note that if $\mathcal{T}=[T]$, then the restricted case is equivalent to the unrestricted case.

\remove{The rest of the paper is organized as follows. 
In Section \ref{section: policy}, we discuss our algorithm, which works based on FTPL algorithm mentioned in \cite{bhattacharjee2020fundamental}.
 Section \ref{sec: Mainresults} contains the main results which describe the performance of our algorithm. In particular, Theorem \ref{theorem:regret_unrestricted} gives an upper bound on the regret with unrestricted switching, Theorem \ref{theorem: advswitch} gives an upper bound on the expected number of switches of our algorithm under unrestricted switching and Theorem \ref{theorem:regret_restricted_switching} gives an upper bound on the regret of our algorithm with a given restricted switching time-slots.
 We conclude with a short discussion about future directions in \ref{sec:conclusion}. Due to the lack of space, the results for the stochastic case and proofs are presented in the \textcolor{red}{Appendix of the full paper.}}

 The rest of the paper is organized as follows. In Section \ref{section: policy}, we discuss our online subset selection algorithm. Section \ref{sec: Mainresults} contains the main results describing the performance of our algorithm. \remove{In particular, Theorem \ref{theorem:regret_unrestricted} provides an upper bound on the regret with unrestricted switching, Theorem \ref{theorem:advswitch} gives an upper bound on the expected number of switches of our algorithm under unrestricted switching, and Theorem \ref{theorem:regret_restricted_switching} provides an upper bound on the regret of our algorithm within given restricted switching time slots.} We include our numerical findings in Section \ref{sec:numerical_experiments} and conclusions and future directions in Section \ref{sec:conclusion}. {Due to spatial limitations, the proofs of our theorems and proposition are provided in the Appendix of the complete version of our paper \cite{nayak2024ontheregret}.}

\section{Our Proposed Algorithm}
\label{section: policy}
\begin{algorithm}
\SetAlgoLined
\footnotesize
\caption{Algorithm for Coded Caching Problem with Adversarial Requests }\label{alg:FTPL}
\begin{algorithmic}[1]
\STATE \textbf{Input:} $M, N, T, \mathbf{s}_0=[1,1\cdots 1], \mathcal{T}, \eta_t=\alpha\sqrt{t} \quad \forall t\in \mathcal{T} $
\STATE Sample $\gamma \sim \mathcal{N}(0,\textit{I}_{N\times N}) $
\FOR{$t \leq T$}
   \STATE \textbf{Placement phase}:
   \STATE Derive $\mathbf{x}_{t-1}$ from $\mathbf{r}_{t-1}$
   \STATE $\mathbf{y}_{t-1} \gets \min\{\mathbf{x}_{t-1}, \mathbb{I}_N\} $
   \STATE $\mathbf{Y}_{t} \gets \mathbf{Y}_{t-1} + \mathbf{y}_{t-1}$
   \IF {$t\in \mathcal{T}$}
      \STATE $\overline{\mathbf{Y}}_{t} = \mathbf{Y}_{t} - \eta_{t}\gamma$
      \STATE \textbf{Subset selection for placement}:
      \STATE $\mathbf{s}_{t} \gets \arg \min_{\mathbf{s}\in\mathcal{S}} \left\langle\left(\mathbf{s} - \frac{M}{N}\mathbb{I}_N\right), \sum_{i = 1}^{t-1}f(\mathbf{x}_i,s) -\overline{\mathbf{Y}}_{t} \right\rangle $
      \STATE Perform cache placement with $\mathbf{s}_t$ according to the placement phase of Section \ref{sec:sysmodel}
         \ELSE
   \STATE $\mathbf{s}_{t}\gets\mathbf{s}_{t-1} $
   \ENDIF
   
   \STATE \textbf{Delivery phase}:
   \STATE Receive $\mathbf{r}_{t}$ 
   \STATE Perform content delivery according to the delivery phase of Section \ref{sec:sysmodel}
\ENDFOR
\end{algorithmic}
\end{algorithm}

In this section, we present our online algorithm, which achieves $\mathcal{O}(\sqrt{T})$ regret. Our algorithm is a variant of the standard FTPL algorithm commonly used in online learning settings, and its pseudocode is provided in Algorithm \ref{alg:FTPL}. Recall that our algorithm has to identify the subset of files $\mathbf{s}_t$ to cache in each time slot $t \in \mathcal{T}$, guided by the history of request patterns $(\mathbf{r}_i)_{i=1}^{t-1}$. A detailed description of our algorithm is given below: As we mentioned earlier, time is divided into slots. During the placement phase,
\begin{itemize}
    \item We first find the vectors $\mathbf{x}_{t-1}$, $\mathbf{y}_{t-1} = \min\{\mathbf{x}_{t-1}, \mathbb{I}_N\} $ and 
    $\mathbf{Y}_{t} = \mathbf{Y}_{t-1} + \mathbf{y}_{t-1}$, based on the request profile $\mathbf{r}_{t-1}$ received in the previous slot $t-1$. (See lines 5 - 7 in Algorithm \ref{alg:FTPL}).
   \item Next, if $t\not\in \mathcal{T}$, we directly start the delivery phase. Otherwise, we update our cache content first before starting the delivery phase. The cumulative rate until slot $t$ that would have been incurred by the cache configuration $\mathbf{s}$  is given by $\left\langle\left(\mathbf{s} - \frac{M}{N}\mathbb{I}_N\right), \sum_{i = 1}^{t-1}f(\mathbf{x}_i,\mathbf{s})-\mathbf{Y}_t \right\rangle + \sum_{i=1}^{t-1}h(\mathbf{x}_i)$. For $t \in \mathcal{T}$, when we are allowed to change the cache configuration, the subset $\mathbf{s}_{t}$ is then determined as the one that minimizes the cumulative rate until $t$ with the perturbed vector  $\overline{\mathbf{Y}}_t=\mathbf{Y}_{t} - \eta_{t}\gamma$ as input. 
   \item After the subset selection, cache placement occurs according to the placement phase described in Section \ref{sec:sysmodel}. (See lines 8-13 in Algorithm \ref{alg:FTPL}).
\end{itemize}
\remove{\begin{itemize}
    \item We first set $\mathbf{y}_{t-1} = \min\{\mathbf{x}_{t-1}, \mathbb{I}_N\} $ and 
    $\mathbf{Y}_{t} = \mathbf{Y}_{t-1} + \mathbf{y}_{t-1}$, based on the request pattern $\mathbf{x}_{t-1}$ received in the previous slot $t-1$. (See lines 5 and 6 in Algorithm \ref{alg:FTPL}).
   \item Next, if $t\not\in \mathcal{T}$, we directly start the delivery phase. Otherwise, we update our cache content first before starting the delivery phase. The cumulative rate until slot $t$ that would have been incurred by the cache configuration $\mathbf{s}$  is given by $\left\langle\left(\mathbf{s} - \frac{M}{N}\mathbb{I}_N\right), \sum_{i = 1}^{t-1}f(\mathbf{x}_i,\mathbf{s})-\mathbf{Y}_t \right\rangle + \sum_{i=1}^{t-1}h(\mathbf{x}_i)$. For $t \in \mathcal{T}$, when we are allowed to change the cache configuration, the subset $\mathbf{s}_{t}$ is then determined as the one that minimizes the cumulative rate until $t$ with the perturbed vector  $\overline{\mathbf{Y}}_t=\mathbf{Y}_{t} - \eta_{t}\gamma$ as input. 
   \item After the subset selection, cache placement occurs according to the placement phase described in Section \ref{sec:sysmodel}. (See lines 7-12 in Algorithm \ref{alg:FTPL}).
\end{itemize}
During the delivery phase,
\begin{itemize}
    \item We first receive the request profile $\mathbf{r}_{t}$. 
    \item After that, based on the user request profile and the cached content, we serve all the users' requested files according to the delivery phase described in Section \ref{sec:sysmodel}. (See lines 14 and 15 in Algorithm \ref{alg:FTPL}).
\end{itemize}}

\remove{In this section, we present our online algorithm, which achieves $\mathcal{O}(\sqrt{T})$ regret under certain conditions on $\mathcal{T}$. Our algorithm aims to identify a subset of files, $\mathbf{s}_t$, to be cached in each time slot $t \in \mathcal{T}$,\ guided by the history of request patterns. Our algorithm is a variant of the standard FTPL algorithm commonly used in online learning settings. The overall subset selection algorithm works as follows
\begin{enumerate}
    \item At $T=0$ we receive inputs $M, N, K$ and sample the Gaussian random vector $\gamma$ from $\mathcal{N}(0,\mathbb{I}_{N\times N})$ as shown in lines 1,2 of Algorithm. \ref{alg:FTPL}.
    \item The vector $\mathbf{y}_t$ is calculated, and $\mathbf{Y}_t$ is updated using $\mathbf{x}_i$ received in slot $t-1$ as detailed in lines 5-6. 
    \item The cumulative rate that would have been incurred by the cache configuration $\mathbf{s}$ until $t$ is given by $\left\langle\left(\mathbf{s} - \frac{M}{N}\mathbb{I}_N\right), \sum_{i = 1}^{t-1}f(\mathbf{x}_i,\mathbf{s})-\mathbf{Y}_t \right\rangle + \sum_{i=1}^{t-1}h(\mathbf{x}_i)$. For $t \in \mathcal{T}$, when we are allowed to change the cache configuration, the subset $\mathbf{s}_{t}$ is then determined as the one that minimizes the cumulative rate until $t$ with the perturbed vector  $\overline{\mathbf{Y}}_t$ as input as per lines 8-10. 
    \item After the subset selection, cache placement occurs according to the guidelines outlined in Section \ref{sec:sysmodel}.
    \item  Subsequently, the delivery phase commences upon receiving requests, as depicted in lines 13-15.
\end{enumerate}

Our proposed algorithm works as }

This is followed by the delivery phase as per Section \ref{sec:sysmodel}. Note that the $h(\mathbf{x}_i)$ terms in the cumulative rate expression are ignored during subset selection (line 11), since they do not depend on the policy parameters $\mathbf{s}$.

\remove{In a single-user caching setting, FTPL is similar to the Least Frequently Used (LFU) algorithm, which monitors the frequency of each file's requests. However, it distinguishes itself by selecting the top $M$ files after introducing a Gaussian perturbation to the cumulative request numbers. The FTPL policy has been extensively studied (see \cite{mukhopadhyay2021online,bhattacharjee2020fundamental,paria2021texttt}) and demonstrated to achieve $\mathcal{O}(\sqrt{T})$ regret in adversarial settings.
However, the rate expression in a single user/uncoded setting is linear in both cache configuration $\mathbf{s}_t$ and cumulative request numbers; as a result of this, the choice of cache configuration is simply the set of $M$ files with the maximum cumulative requests after perturbation. Since the expected rate expression in the coded caching setting is a non-linear function of the request pattern $\mathbf{x}_t$ as well as the cache configuration $(\mathbf{s}_t)$, adding the perturbation to request pattern $\mathbf{x}_t$ directly makes the regret analysis quite difficult.}

\remove{In a single-user caching scenario, FTPL resembles the Least Frequently Used (LFU) algorithm, which tracks the frequency of each file's requests. However, FTPL distinguishes itself by selecting the top $M$ files after introducing a Gaussian perturbation to the cumulative request numbers. }
In a single-user caching scenario, FTPL tracks the frequency of each file's requests and selects the top $M$ files after introducing a Gaussian perturbation to the cumulative request numbers.
Extensively studied in prior works \cite{mukhopadhyay2021online,zarin2022regret}, the FTPL policy has proven to achieve $\mathcal{O}(\sqrt{T})$ regret in adversarial settings. {In a single-user setting, the rate expression is linear in both cache configuration $\mathbf{s}_t$ and cumulative request numbers.} Consequently, the cache configuration choice simply involves selecting the set of $M$ files with the maximum cumulative requests after perturbation. However, in the coded caching setting, the expected rate expression becomes a non-linear function of the request pattern $\mathbf{x}_t$ and the cache configuration $\mathbf{s}_t$. Introducing perturbation directly to the request pattern $\mathbf{x}_t$ complicates the regret analysis significantly.

Nevertheless, it is noteworthy that the expected cumulative rate that would be incurred for any cache configuration $\mathbf{s}$ until time $t$ is given by $\sum_{i = 1}^{t-1} K(\mathbf{s},\mathbf{x}_i) = \left\langle\left(\mathbf{s}-\frac{M}{N}\mathbb{I}_N\right), \sum_{i = 1}^{t-1}f(\mathbf{s}_i,\mathbf{x})-\mathbf{Y}_t\right\rangle +  \sum_{i = 1}^{t-1}h(\mathbf{x}_i),$ and it is linear in $\mathbf{Y}_t$. Here, $\mathbf{Y}_t = \sum_{i=1}^{t-1}\mathbf{y}_i$ is a function of the requests. Our Algorithm relies on perturbing $\mathbf{Y}_t$ for subset selection at $t\in\mathcal{T}$. The performance of our proposed policy under various scenarios is evaluated in Section \ref{sec: Mainresults}.

\begin{figure*} 
\centering
\subfloat
{\includegraphics[width=0.53\columnwidth]{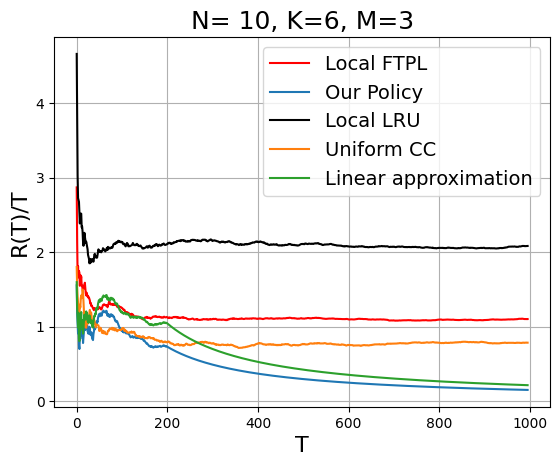}}
\hfil
\label{fig:figure1}
\subfloat
{\includegraphics[width=0.53\columnwidth]{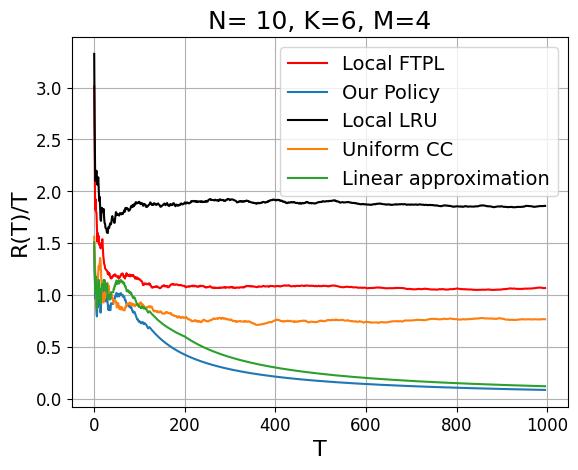}}
\hfil
\subfloat
{\includegraphics[width=0.53\columnwidth]{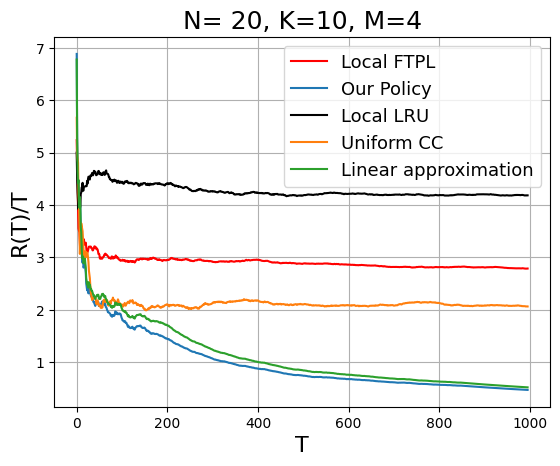}}
\caption{\sl We compare the performance of our proposed policy in Algorithm \ref{alg:FTPL} against the benchmark policies under different system parameters. The plots display the average regret per time step $R(T)/T$ against horizon $T$. The first figure compares the performance of benchmark policies against our policy for $N=10$ files, $K=6$ users, and cache size $M=3$. In the next figure, we increase the cache size to $M=4$. The last figure for the case $N=20$, $K=10$, and $M=4$. Our policy outperforms the benchmarks and has a sub linear regret which matches our theoretical findings.
}
\vspace{-12pt}
\label{fig:figure2}
\end{figure*}

\section{Main Results}
\label{sec: Mainresults}
Now, we discuss the performance of our algorithm under the adversarial requests setting. We evaluate the performance of our algorithm in two different scenarios. The first scenario is unrestricted switching, where there are no restrictions on switching cache configurations; i.e., $\mathcal{T} = \{1,2\cdots T\}$, we are allowed to switch in every time slot. The second scenario is restricted switching, where we can only switch in a set of pre-specified time slots.

\subsubsection{Unrestricted Switching} 
\remove{\color{red}Let constant $r_{\max}$ be the maximum length of the server transmission (coded + uncoded)over the set of all request pattern and cache configuration pairs, respectively. Note that it can be upper bounded by $K$ as the maximum number of distinct requests per time slot will be $K$.} Let $|\mathcal{S}|$ be the cardinality of the set of feasible cache configurations. The following result gives an upper bound on the adversarial regret of Algorithm \ref{alg:FTPL} with no restrictions on switching.
\remove{\begin{theorem} \label{theorem:regret_unrestricted}
    The regret incurred by the proposed online policy (Algorithm \ref{alg:FTPL}) in the adversarial setting w.r.t. the oracle performance described in \eqref{eqn:oracle_performace_general} for a horizon $T$ can be upper bounded by
\begin{multline}
        R_{(\eta, \text{UR})}^{A}(T) \leq   r_{\max}^C  +\eta_T\mathcal{G}_{\max}(\gamma)+
         \\ \left(\frac{K^2}{\sqrt{2\pi}} +\frac{3r_{\max}r_{\max}^C|\mathcal{S}|}{2\sqrt{\pi}}\right)\sum_{t=1}^T\frac{1}{\eta_t} = \mathcal{O}(\sqrt{T}) 
\end{multline}
where \\
$\eta_t = \alpha\sqrt{t}$ and $\mathcal{G}_{\max}(\gamma)$ is the maximum Gaussian width given by
\begin{equation}
    \max_{a \in\mathcal{A}}\mathbb{E}_{\gamma}[\langle a,\gamma\rangle] 
\end{equation}
Here 
\[\mathcal{A} = \left\{a=[a_1,a_2,...,a_N]^T|a_i \in \{0,1\};\langle a, \mathbb{I}_N\rangle\geq M\right\} \]
and $\gamma \sim \mathcal{N}(0,\textit{I}_{N\times N})$.
This can be upper bounded as $\max_{M\leq  m\leq N}m\sqrt{\log(\frac{N}{m})}.$
\end{theorem}

\begin{theorem} \label{theorem:regret_unrestricted}
For a coded caching problem with $N$ files, $K$ users/caches and cache size $MF$ bits, let $ R_{ \text{UR}}(T)$  be the Algorithm \ref{alg:FTPL}'s adversarial regret with unrestricted switching. Then, 
\begin{align*}
        R_{(\mathbf{\eta}, \text{UR})}^{A}(T) &\leq   r_{\max}^C  +\eta_T\mathcal{G}_{\max}(\gamma)+
         \\ &\left(\frac{K^2}{\sqrt{2\pi}} +\frac{3r_{\max}r_{\max}^C|\mathcal{S}|}{2\sqrt{\pi}}\right)\sum_{t=1}^T\frac{1}{\eta_t} = \mathcal{O}(\sqrt{T}) 
\end{align*}
where $\mathcal{G}_{\max}(\gamma)$ is the maximum Gaussian width given by
    $\max_{a \in\mathcal{A}}\mathbb{E}_{\gamma}[\langle a,\gamma\rangle]$ 
with
\[\mathcal{A} = \left\{a=[a_1,a_2,...,a_N]^T|a_i \in \{0,1\};\langle a, \mathbb{I}_N\rangle\geq M\right\} \]
and $\gamma \sim \mathcal{N}(0,\textit{I}_{N\times N})$.
$\mathcal{G}_{\max}(\gamma)$ can be upper bounded as $\max_{M\leq  m\leq N}m\sqrt{\log(\frac{N}{m})}.$
\end{theorem}}
\begin{theorem} \label{theorem:regret_unrestricted}
For a coded caching problem with $N$ files, $K$ users/caches, and cache size $MF$ bits, let $R_{ \text{UR}}(T)$ be the adversarial regret of Algorithm \ref{alg:FTPL} with unrestricted switching. Then, 
\begin{multline*}
    R_{ \text{UR}}(T) \leq c_1 + c_2\sqrt{T} 
    +\frac{K^2(2+3|\mathcal{S}|)}{2\sqrt{2\pi}\alpha}\sum_{t=1}^T \frac{1}{\sqrt{t}}=\mathcal{O}(\sqrt{T})
\end{multline*}
where $c_1$ and $c_2$ are constants.
\end{theorem}
\remove{As we mentioned in Section \ref{sec:sysmodel}, an algorithm that fetches a large number of files into the caches each time to change the cache configuration is not ideal as fetching files into the caches causes latency and consumes bandwidth.  Hence, in Theorem \ref{theorem:advswitch}, we present an upper bound on the expected number of switches incurred by Algorithm \ref{alg:FTPL} under unrestricted switching.}
In Theorem \ref{theorem:advswitch}, we present an upper bound on the expected number of switches incurred by Algorithm \ref{alg:FTPL} under unrestricted switching.
\begin{theorem}
\label{theorem:advswitch}
For a coded caching problem with $N$ files, $K$ users/caches, and cache size $MF$ bits, let $C_{\text{UR} }(T)$ denote
the expected number of switches in cache configuration until time $T$ for Algorithm \ref{alg:FTPL} with unrestricted switching. Then
\begin{equation}
   C_{\text{UR} }(T) \leq \frac{3K(|\mathcal{S}|-1)}{2\sqrt{2\pi}\alpha} \sum_{t=1}^T \frac{1}{\sqrt{t}}=\mathcal{O}(\sqrt{T}). \nonumber 
\end{equation}
\end{theorem}
\begin{remark}
    From Theorems \ref{theorem:regret_unrestricted} and \ref{theorem:advswitch}, we conclude that Algorithm \ref{alg:FTPL} with unrestricted switching achieves $\mathcal{O}(\sqrt{T})$ adversarial regret by just using $\mathcal{O}(\sqrt{T})$ expected cache configuration switches.
\end{remark}
\subsubsection{Restricted Switching: } 
In this scenario, the cache content can only be changed in restricted time slots given by the set $\mathcal{T}$; i.e., we are not allowed to change cache configuration outside these time slots. Let $l_k \triangleq t_k-t_{k-1}$ define the time gap between the $(k-1)^{th}$ and $k^{th}$ switching slot, and $L \triangleq |\mathcal{T}|$ define the maximum number of allowed switches. Recall that by convention, $t_0 = 0$ and $t_L = T$, and we have $\sum_{k=1}^L l_k = T$.  The following result gives an upper bound on the adversarial regret of Algorithm \ref{alg:FTPL} when cache updates are restricted to time slots in the set $\mathcal{T}$.
\remove{\begin{theorem}\label{theorem:regret_restricted_switching}
Let $R_{(\eta, \text{UR})}^{A}(T)$ be the regret incurred by the policy (algorithm \ref{alg:FTPL}) in the adversarial unrestricted switching case with a horizon $T$. Then, the regret incurred by this policy in the restricted switching scenario with a learning rate $\alpha\sqrt{t}$ is upper-bounded as
\begin{equation}
R_{(\eta, \text{R})}^{A}(T) \leq R_{(\eta, \text{UR})}^{A}(T) +  \sum_{k=1}^{L}\frac{3r_{\max}^2(|\mathcal{S}|-1)l_k(l_k-1)}{4\alpha\sqrt{\pi}\sqrt{\sum_{i=1}^{k-1} l_i+1}} 
\end{equation}
Here $L$ is the number of switching slots until $T$.
\end{theorem}}
\begin{theorem}\label{theorem:regret_restricted_switching}
For a coded caching problem with $N$ files, $K$ users/caches, and cache size $MF$ bits, let $R_{ \text{R}}^{\mathcal{T}}(T)$ be the adversarial regret of Algorithm \ref{alg:FTPL} under restricted switching with switching slots $\mathcal{T}$. Then,
\begin{equation}
R_{\text{R}}^{\mathcal{T}}(T) \leq R_{\text{UR}}(T) +  \sum_{k=1}^{L}\frac{3K^2(|\mathcal{S}|-1)l_k(l_k-1)}{4\alpha\sqrt{\pi}\sqrt{\sum_{i=1}^{k-1} l_i+1}}. \nonumber
\end{equation}
\end{theorem}
\begin{remark}
If $\mathcal{T}=[T]$, then we get back the unrestricted switching setting. In this case, the upper bound collapses to $R_{ \text{UR}}(T)$, and the second term in RHS becomes zero. Also, for a fixed intermittent switching period $l_k = l$ $\forall k$, i.e., we have $\left\lfloor\ T/l\right\rfloor +1$ total switching slots, and the second term in the upper bound grows as $\mathcal{O}(l\sqrt{T})$.
\end{remark}

\section{Numerical Experiments}\label{sec:numerical_experiments}
In this section, we compare the performance of our policy against the following benchmark policies.

1) \textbf{Local caching with FTPL}: During the placement phase we use the FTPL algorithm \cite{mukhopadhyay2021online} independently at each user on the requests until time $t-1$ to select the files to be cached at time $t$ in their respective local caches. In essence, the policy stores the $M$ most popular files locally based on the requests until time $t-1$ at each user after Gaussian perturbation. Thus, the user's cache will be representative of their preferences.
\remove{{\color{red}Let $z_t^k$ be an $N$-dimensional vector with an entry "1" for every file stored in the cache at user $k$ and "0" otherwise at time $t$. For a request profile $r_k^t$, again a $N$-dimensional vector with an entry "1" for the file requested by user $k$ at $t$ and "0" otherwise, the rate incurred at time $t$ will be $\sum_{n=1}^{N}\min\left\{\sum_{k=1}^K\langle 1-z_k^t,r_k^t\rangle, 1 \right\}$. The minimum comes from only one transmission per file being required in a broadcast medium, even when multiple users request the same file.}}
 
2) \textbf{Local LRU}: During the placement phase, stores the last $M$ ("Least Recently Used") requested files by user $k$ in its cache at any point $t$. 

For both of these policies, during the delivery phase, requests for files that are present in their respective user caches are served directly by the cache, while the remaining requests are served by the server through the broadcast channel. Furthermore, only one transmission is needed if multiple users request the same file.

3) \textbf{Uniform Coded Caching}: This policy stores equal fractions of all the $N$ files at each user ($\mathbf{s}_t=\mathbb{I}_N$ $\forall t$) and uses the coded delivery method described in section \ref{sec:sysmodel}.

4) \textbf{Linear approximation}: This policy uses a linear approximation of the rate expression to compute $\mathbf{s}_t$ and uses it to perform coded delivery. The scheme does not always achieve a sublinear regret. More details in Appendix \ref{app:linapp}

All the simulations were performed on the Movielens 1M dataset \cite{ML1m_paper} containing $\sim$ 1 million ratings from 6040 users on 3706 movies. We restricted our analysis to a subset of files with a significant number of requests (>1000) and randomly selected $N$ files from this subset. To ensure an adequate number of requests per user from the restricted file set, we combined multiple users into "virtual" users.

\remove{LFU (Local FTPL without perturbations) performs similar to the local FTPL in the Movielens simulations above. However, it is known that under alternating requests at local users with a period $M$, e.g., M=2, $\{1,2,2,1,1,2,2\cdots\}$, LFU will perform badly with a hit rate 0, and a cumulative rate that grows as $\mathcal{O}(T)$. \cite{mukhopadhyay2021online},\cite{zarin2022regret}.}

\section{Conclusions and Future Work}\label{sec:conclusion}

Our work focuses on the widely studied coded caching problem using the lens of online learning under adversarial settings. We propose an algorithm and show that our policy achieves $\mathcal{O}(\sqrt{T})$ adversarial regret. We also consider the issue of switching cost by providing an upper bound on the expected number of switches of our algorithm under unrestricted switching and giving an upper bound on the adversarial regret with restricted switching. \remove{Additionally, we support our theoretical findings with numerical results.}


There are several potential avenues for future work, starting from deriving universal lower bounds on Regret. Another interesting problem is proposing algorithms for the case where switching budgets are fixed. Unlike our restricted switching case with given switching slots, fixed switching budgets would require the policy to decide when to switch the cache configuration within budget constraints. One more interesting problem is addressing the issue of online learning with expert advice within the context of coded caching.

\clearpage
\bibliographystyle{IEEEtran}
\bibliography{references}

\newpage
\onecolumn
\title{Appendix}

\section{Appendix}\label{sec:app}
\subsection{Proof of Proposition \ref{proposition:expected_rate}}
\label{app: rateexp}
\begin{proof}
    Recall from Section \ref{sec:sysmodel} that we perform both coded and uncoded transmissions during the delivery phase. Let $R_{UC}(\mathbf{s}_t,\mathbf{x}_t)$ and $R_C(\mathbf{s}_t,\mathbf{x}_t)$ denote the expected rate of the uncoded and coded transmissions associated with a cache configuration $\mathbf{s}_t$ and the request pattern $\mathbf{x}_t$, respectively. Therefore, $K(\mathbf{s}_t,\mathbf{x}_t)=R_{UC}(\mathbf{s}_t,\mathbf{x}_t)+R_C(\mathbf{s}_t,\mathbf{x}_t)$. One can calculate $R_{UC}(\mathbf{s}_t,\mathbf{x}_t)$ and $R_C(\mathbf{s}_t,\mathbf{x}_t)$ separately as follows:

\subsubsection*{ $R_{UC}$ calculation}
If an \textit{unstored file} is requested by at least one of the users, then it is directly broadcast by the server. Therefore, the expected uncoded transmission rate $R_{UC}(\mathbf{s}_t,\mathbf{x}_t)$ is given by 
\begin{equation}
    R_{UC}(\mathbf{s}_t,\mathbf{x}_t) = \langle (\mathbb{I}_N - \mathbf{s}_t), \min\{\mathbf{x}_t, \mathbb{I}_N\} \rangle = \langle (\mathbb{I}_N - \mathbf{s}_t), \mathbf{y}_t \rangle. \label{eqn:uncoded_transmission_size}
\end{equation}

\subsubsection*{$R_{C}$ calculation}
Let $U_1$ denote the set of users that request files from the cached set $\mathbf{s}_t$, then $|U_1| = \langle \mathbf{x}_t, \mathbf{s}_t \rangle$. Recall from Section \ref{sec:sysmodel} that, during the coded transmission,  
for every subset $u \in U_1$ with $|u| \neq 0$, transmit $\bigoplus_{k \in u} V_{k, u\setminus\{k\}}$.
Here, $V_{k, u\setminus\{k\}}$ denotes all the bits that are requested by user $k \in u$, are present in the cache of all users in $u\setminus\{k\}$, and that are not stored in the caches of any other user in $U_1\setminus u$. 
Therefore, the expected length of the coded transmission $R_C(\mathbf{s}_t,\mathbf{x}_t)$ is equal to the expected sum of lengths of all these messages $\bigoplus_{k\in u} V_{k,u\setminus\{k\}}$ for all $u\subseteq U_1$, where the length of a message $\bigoplus_{k\in u} V_{k,u\setminus\{k\}}$ is $|\bigoplus_{k\in u} V_{k,u\setminus\{k\}}| = \max_{k\in u} | V_{k,u\setminus\{k\}}|$.

Since the users independently select $\frac{M}{\langle \mathbf{s}_t, \mathbb{I}_N \rangle}$ fraction of each \textit{stored file}, the probability of a particular bit of a \textit{stored file} being in the cache of a particular user is $\frac{M}{\langle \mathbf{s}_t, \mathbb{I}_N \rangle}$. Therefore, for large enough $F$, we have
\begin{align}
    \max_{k\in u} | V_{k,u\setminus\{k\}}| &= F\left(\frac{M}{\langle \mathbf{s}_t,\mathbb{I}_N \rangle} \right)^{|u|-1} \left(1-\frac{M}{\langle \mathbf{s}_t,\mathbb{I}_N \rangle} \right)^{\langle \mathbf{x}_t,\mathbf{s}_t \rangle-|u|+1} + o(F) \nonumber \\
    & \approx F\left(\frac{M}{\langle \mathbf{s}_t,\mathbb{I}_N \rangle} \right)^{|u|-1} \left(1-\frac{M}{\langle \mathbf{s}_t,\mathbb{I}_N \rangle} \right)^{\langle \mathbf{x}_t,\mathbf{s}_t \rangle-|u|+1}. \nonumber 
\end{align}
Now for each $|u| \in \{1, 2, \ldots, \langle \mathbf{x}_t, \mathbf{s}_t \rangle\}$, we have ${\langle \mathbf{x}_t, \mathbf{s}_t \rangle \choose |u|}$ subsets of size $|u|$. Therefore, the coded transmission message size $R_C(\mathbf{s}_t,\mathbf{x}_t) \cdot F$ is
\begin{align}
       R_C(\mathbf{s}_t,\mathbf{x}_t)\cdot F  & = F\sum_{|u|=1}^{\langle \mathbf{x}_t, \mathbf{s}_t\rangle} {\langle \mathbf{x}_t,\mathbf{s}_t\rangle \choose |u|}\left(\frac{M}{\langle \mathbf{s}_t,\mathbb{I}_N \rangle} \right)^{|u|-1} \left(1-\frac{M}{\langle \mathbf{s}_t,\mathbb{I}_N \rangle} \right)^{\langle \mathbf{x}_t,\mathbf{s}_t \rangle-|u|+1} \nonumber \\
        & = F\left(\frac{\langle \mathbf{s}_t,\mathbb{I}_N\rangle}{M}-1\right)\left(1-\left(1-\frac{M}{\langle \mathbf{s}_t, \mathbb{I}_N\rangle}\right)^{\langle \mathbf{x}_t,\mathbf{s}_t\rangle}\right). \label{eqn:coded_transmission_size}
\end{align}

From \eqref{eqn:uncoded_transmission_size} and \eqref{eqn:coded_transmission_size}, we have \begin{equation}
K(\mathbf{s}_t,\mathbf{x}_t)=R_{UC}(\mathbf{s}_t,\mathbf{x}_t)+R_{C}(\mathbf{s}_t,\mathbf{x}_t)=\langle (\mathbb{I}_N-\mathbf{s}_t),\mathbf{y}_t\rangle+\left(\frac{\langle \mathbf{s}_t,\mathbb{I}_N\rangle}{M}-1\right)\left(1-\left(1-\frac{M}{\langle \mathbf{s}_t, \mathbb{I}_N\rangle}\right)^{\langle \mathbf{x}_t,\mathbf{s}_t\rangle}\right)
\end{equation}
\end{proof}

\subsection{Proof of Theorem \ref{theorem:regret_unrestricted}}
\label{app: advregret}
We will use the following lemmas to prove Theorem \ref{theorem:regret_unrestricted}.
\begin{lemma}\label{lemma:rewrite}
    For a coded caching problem with the given cache configuration $\mathbf{s}_t$ and the request vector $\mathbf{x}_t$, the placement and delivery policies discussed in Section \ref{sec:sysmodel} transmits a message of expected size $K(\mathbf{s}_t,\mathbf{x}_t)$ given by
    \begin{equation}
    K(\mathbf{s}_t,\mathbf{x}_t) = \left\langle\left(\mathbf{s}_t-\frac{M}{N}\mathbb{I}_N\right), f(\mathbf{x}_t,\mathbf{s}_t)-\mathbf{y}_t\right\rangle +  h(\mathbf{x}_t),
\end{equation}
where $f(\mathbf{x}_t,\mathbf{s}_t) = \frac{1}{M}\left(1-\left(1-\frac{M}{\langle \mathbf{s}_t, \mathbb{I}_N\rangle}\right)^{\langle \mathbf{x}_t,\mathbf{s}_t\rangle}\right)\mathbb{I}_N $ and $h(\mathbf{x}_t) = \left(1- \frac{M}{N}\right)\langle \mathbf{y}_t, \mathbb{I}_N\rangle$.
\end{lemma}
\begin{proof}
    Using the fact that $\langle \mathbb{I}_N, \mathbb{I}_N\rangle = N $, we have
\begin{align*}
    R_{C}(\mathbf{s}_t,\mathbf{x}_t)&=\left(\frac{\langle \mathbf{s}_t,\mathbb{I}_N\rangle}{M}-1\right)\left(1-\left(1-\frac{M}{\langle \mathbf{s}_t, \mathbb{I}_N\rangle}\right)^{\langle \mathbf{x}_t,\mathbf{s}_t\rangle}\right)\\
    & = \left\langle\mathbf{s}_t, \frac{1}{M}\left(1-\left(1-\frac{M}{\langle \mathbf{s}_t, \mathbb{I}_N\rangle}\right)^{\langle \mathbf{x}_t,\mathbf{s}_t\rangle}\right) \mathbb{I}_N\right\rangle - \left\langle\frac{M}{N}\mathbb{I}_N, \frac{1}{M}\left(1-\left(1-\frac{M}{\langle \mathbf{s}_t, \mathbb{I}_N\rangle}\right)^{\langle \mathbf{x}_t,\mathbf{s}_t\rangle}\right) \mathbb{I}_N\right\rangle\\
    & = \left\langle \left(\mathbf{s}_t - \frac{M}{N}\mathbb{I}_N\right), f(\mathbf{x}_t,\mathbf{s}_t)\right\rangle
\end{align*}
and
\begin{align*}
    R_{UC}(\mathbf{s}_t,\mathbf{x}_t) = \langle (\mathbb{I}_N - \mathbf{s}_t), \mathbf{y}_t \rangle = \left\langle \left(\mathbf{s}_t - \frac{M}{N}\mathbb{I}_N\right), -\mathbf{y}_t \right\rangle + \left\langle\left(1-\frac{M}{N}\right)\mathbb{I}_N, \mathbf{y}_t \right\rangle.
\end{align*}
Therefore,
\begin{equation*}
    K(\mathbf{s}_t,\mathbf{x}_t) = R_{C}(\mathbf{s}_t,\mathbf{x}_t)+ R_{UC}(\mathbf{s}_t,\mathbf{x}_t) = \left\langle\left(\mathbf{s}_t-\frac{M}{N}\mathbb{I}_N\right), f(\mathbf{x}_t,\mathbf{s}_t)-\mathbf{y}_t\right\rangle +  h(\mathbf{x}_t).
\end{equation*}
\end{proof}
 
Let the constant $r_{\max}^C$ denotes the maximum size of the coded transmission message over the set of all request pattern and cache configuration pairs, i.e.,
\begin{equation*}
    r^C_{\max} = \max_{\substack{\mathbf{x}\in X\\ \mathbf{s}\in \mathcal{S}}} R_C(\mathbf{s},\mathbf{x}),
\end{equation*}
and the constant $r_{\max}$ denotes the maximum size of the server's transmission message over the set of all request pattern and cache configuration pairs, i.e.,
\begin{equation*}
    r_{\max} = \max_{\substack{\mathbf{x}\in X\\ \mathbf{s}\in \mathcal{S}}} K(\mathbf{s},\mathbf{x}).
\end{equation*}
 Note that both $r_{\max}$ and $r_{\max}^C$ can be upper bounded by $K$ since the maximum number of unique requests in a slot is $K$. For a random vector $\gamma =[\gamma_1,\gamma_2,...,\gamma_N]^T \sim \mathcal{N}(0,\textit{I}_{N\times N})$, and a cache configuration $\mathbf{a}\in \mathcal{S}$, we define a constant  $\mathcal{G}_{\max}(\gamma)$  as
    \[\mathcal{G}_{\max}(\gamma)=\mathbb{E}_{\gamma}\left[\max_{a \in\mathcal{S}}\langle a,\gamma\rangle\right] = -\mathbb{E}_{\gamma}\left[\min_{a \in\mathcal{S}}\langle a,\gamma\rangle\right].\]
 Note that $\mathcal{G}_{\max}(\gamma)$ can be easily upper bounded as $\mathcal{G}_{\max}(\gamma)=\mathbb{E}_{\gamma}\left[\max\limits_{a \in\mathcal{S}}\langle a,\gamma\rangle\right] \leq \mathbb{E}_{\gamma}\left[\max\limits_{a \in\{0,1\}^N}\langle a,\gamma\rangle\right] = \sum\limits_{i=1}^N\gamma_i\mathbb{I}(\gamma_i\geq 0) \leq  \frac{N}{\sqrt{2\pi}}$. 

\begin{lemma}
    For a coded caching problem with $N$ files, $K$ users/caches, and cache size $MF$ bits, let $R_{\text{UR}}(T)$ be the regret under adversarial requests incurred by Algorithm \ref{alg:FTPL} under unrestricted switching with  $\eta_t = \alpha\sqrt{t}$. Then, \begin{multline}
        R_{\text{UR}}(T) \leq  \left(\frac{3r_{\max}r_{\max}^C(|\mathcal{S}|-1)}{2\sqrt{2\pi}\alpha}+\frac{\max\left\{\frac{M}{N},\left(1-\frac{M}{N}\right)\right\}K^2}{\sqrt{2\pi}\alpha}\right)\sum_{t=1}^T\frac{1}{\sqrt{t}} +\alpha\sqrt{T}\mathcal{G}_{\max}(\gamma)
        + \eta_1\mathcal{G}_{\max}(\gamma) + r_{\max}^C \nonumber
\end{multline}
\end{lemma}

\begin{proof}
From Proposition \ref{proposition:expected_rate}, we know that, for a given cache configuration $\mathbf{s}_t$ and a request pattern $\mathbf{x}_t$, the expected transmission size $K(\mathbf{s}_t,\mathbf{x}_t)$ is 
\[
K(\mathbf{s}_t,\mathbf{x}_t)=R_{UC}+R_{UC}=\langle (\mathbb{I}_N-\mathbf{s}_t),\mathbf{y}_t\rangle+\left(\frac{\langle \mathbf{s}_t,\mathbb{I}_N\rangle}{M}-1\right)\left(1-\left(1-\frac{M}{\langle \mathbf{s}_t, \mathbb{I}_N\rangle}\right)^{\langle \mathbf{x}_t,\mathbf{s}_t\rangle}\right).
\]
 From Lemma \ref{lemma:rewrite}, $K(\mathbf{s}_t,\mathbf{x}_t)$ can be rewritten as
\begin{equation}
K(\mathbf{s}_t,\mathbf{x}_t)=\left\langle\left(\mathbf{s}_t - \frac{M}{N}\mathbb{I}_N\right),\left(f(\mathbf{x}_t,\mathbf{s}_t) - \mathbf{y}_t\right) \right\rangle + h(\mathbf{x}_t),
\end{equation}
where,
$f(\mathbf{x}_t,\mathbf{s}_t) = \frac{1}{M}\left(1-\left(1-\frac{M}{\langle \mathbf{s}_t, \mathbb{I}\rangle}\right)^{\langle \mathbf{x}_t,\mathbf{s}_t\rangle}\right) \mathbb{I}_N $ and
$h(\mathbf{x}_t) = \left(1- \frac{M}{N}\right)\langle \mathbf{y}_t, \mathbb{I}_N\rangle$. 

For a request pattern $(\mathbf{x}_t)_{t=1}^T$, let $R_{\text{UR}}((\mathbf{x}_t)_{t=1}^T,T)$ denote the regret incurred by Algorithm \ref{alg:FTPL}. From equations \eqref{eqn:algortim_regret_general}, we have 
\begin{align}  
    R_{\text{UR}}((\mathbf{x}_t)_{t=1}^T,T) &=K^{\pi}(T) - K^{\text{o}}(T)\\
    &=\sum_{t=1}^T\mathbb{E}[K(\mathbf{s}_t,\mathbf{x}_t)]-\min_{\mathbf{s}\in \mathcal{S}}\sum_{t=1}^T\mathbb{E}[K(\mathbf{s},\mathbf{x}_t)] \quad (\text{from equations } \eqref{eqn:algortim_performace_general} \text{ and } \eqref{eqn:oracle_performace_general}) \nonumber\\
    &= \underbrace{\sum_{t=1}^{T}\mathbb{E}_{\gamma}\left[\left\langle\left(\mathbf{s}_t - \frac{M}{N}\mathbb{I}_N\right),\left(f(\mathbf{x}_t,\mathbf{s}_t) - \mathbf{y}_t\right)\right\rangle + h(\mathbf{x}_t)\right]}_{\text{Policy Cumulative Rate}} - \nonumber \\
    &\hspace{2.5in}\underbrace{\min_{s\in\mathcal{S}}\sum_{t=1}^{T} 
    \left[\left\langle\left(\mathbf{s} - \frac{M}{N}\mathbb{I}_N\right),\left(f(\mathbf{x}_t,\mathbf{s}) - y_t\right)\right\rangle + h(\mathbf{x}_t)\right]}_{\text{Oracle Cumulative Rate}} \nonumber \\
     &= \sum_{t=1}^{T}\mathbb{E}_{\gamma}\left[\left\langle\left(\mathbf{s}_t - \frac{M}{N}\mathbb{I}_N\right),\left(f(\mathbf{x}_t,\mathbf{s}_t) - y_t\right)\right\rangle\right] -  \min_{s\in\mathcal{S}}\left\langle\left(\mathbf{s} - \frac{M}{N}\mathbb{I}_N\right), \sum_{t=1}^{T}\left(f(\mathbf{x}_t,\mathbf{s}) - y_t\right)\right\rangle \label{eq:7}
\end{align}
The last equality comes from canceling $h(\mathbf{x}_t)$ terms on both sides, which are independent of the cache configurations. 

Recall from Section \ref{section: policy} that, we define the vector $\mathbf{Y}_t$ as $\mathbf{Y}_t \triangleq \sum_{i=1}^{t-1}\mathbf{y}_i$. Therefore, the term  $\mathbb{E}_{\gamma}\left[\left\langle \left(\mathbf{s}_t-\frac{M}{N}\mathbb{I}_N\right),(f(\mathbf{x}_t,\mathbf{s}_t) - \mathbf{y}_t)\right\rangle\right]$ in \eqref{eq:7} can be written as 
\begin{align}
    \mathbb{E}_{\gamma}\left[\left\langle \left(\mathbf{s}_t-\frac{M}{N}\mathbb{I}_N\right),(f(\mathbf{x}_t,\mathbf{s}_t) - \mathbf{y}_t)\right\rangle\right]
    &= \mathbb{E}_{\gamma}\left[\left\langle \left(\mathbf{s}_t-\frac{M}{N}\mathbb{I}_N\right),f(\mathbf{x}_t,\mathbf{s}_t) - (\mathbf{Y}_{t+1}-\mathbf{Y}_{t})\right\rangle\right] \nonumber \\
    &=\mathbb{E}_{\gamma}\left[\left\langle \left(\mathbf{s}_t-\frac{M}{N}\mathbb{I}_N\right),f(\mathbf{x}_t,\mathbf{s}_t)\right\rangle\right] - \mathbb{E}_{\gamma}\left[\left\langle \left(\mathbf{s}_t-\frac{M}{N}\mathbb{I}_N\right),(\mathbf{Y}_{t+1}-\mathbf{Y}_{t})\right\rangle\right].\label{eq:regretff}
\end{align}

{Let $\mathbf{W} \triangleq (\mathbf{w}_i)_{i=1}^{t-1}$ be the collection of $t-1$ $\mathbf{w}_i$ vectors of dimension $N$ and $\mathbf{Z}$ be a vector of dimension $N$, then, we define a potential function $\phi_t(\mathbf{W},\mathbf{Z})$ as \begin{equation}
    \phi_t(\mathbf{W},\mathbf{Z}) = \mathbb{E}_{\gamma}\left[\min_{\mathbf{s} \in \mathcal{S}}\left\langle\left(\mathbf{s} - \frac{M}{N}\mathbb{I}_N\right),\sum_{i=1}^{t-1}f(\mathbf{w}_i,\mathbf{s}) -\mathbf{Z} + \eta_t \gamma \right\rangle\right].
\label{eq:potential}
\end{equation}
Similar to Equation (5) in \cite{abernethy2014online}, for a given history of request patterns $X_t = (\mathbf{x}_i)_{i=1}^{t-1}$ until time $t$ and  $\mathbf{Y}_t = \sum_{i=1}^{t-1}\mathbf{y}_i$, we have the partial derivative of the potential function as 
\begin{equation}
    \nabla_\mathbf{Y} \phi_t(\mathbf{X}_t,\mathbf{Y}_t) = -\mathbb{E}_{\gamma}\left[\mathbf{s}_t-\frac{M}{N}\mathbb{I}_N\right].
\end{equation}}
Using the Taylor series expansion and for some $\Tilde{\mathbf{Y}_t} =\mathbf{Y}_t+\theta \mathbf{y}_t$ for some $\theta \in [0,1]$, we have
\begin{align}
-\mathbb{E}_{\gamma}\left[\left\langle \left(\mathbf{s}_t-\frac{M}{N}\mathbb{I}_N\right),(\mathbf{Y}_{t+1}-\mathbf{Y}_{t})\right\rangle\right]
& = \langle\nabla_\mathbf{Y}\phi_t(\mathbf{X}_t,\mathbf{Y}_t),(\mathbf{Y}_{t+1}-\mathbf{Y}_t)\rangle \nonumber \\
& = \phi_t(\mathbf{X}_t,\mathbf{Y}_{t+1})- \phi_t(\mathbf{X}_t,\mathbf{Y}_t) - \frac{1}{2}\left\langle \mathbf{y}_t, \nabla_\mathbf{Y}^2 \phi_t(\mathbf{X}_t, \Tilde{\mathbf{Y}}_t) \mathbf{y}_t\right\rangle \label{eqn:Taylor}
\end{align}

Therefore, 
\begin{align}
    &\sum_{t=1}^T\mathbb{E}_{\gamma}\left[\left\langle \left(\mathbf{s}_t-\frac{M}{N}\mathbb{I}_N\right),(f(\mathbf{x}_t,\mathbf{s}_t) - \mathbf{y}_t)\right\rangle\right] \nonumber\\    &=\sum_{t=1}^T\mathbb{E}_{\gamma}\left[\left\langle \left(\mathbf{s}_t-\frac{M}{N}\mathbb{I}_N\right),f(\mathbf{x}_t,\mathbf{s}_t)\right\rangle\right] - \sum_{t=1}^T\mathbb{E}_{\gamma}\left[\left\langle \left(\mathbf{s}_t-\frac{M}{N}\mathbb{I}_N\right),(\mathbf{Y}_{t+1}-\mathbf{Y}_{t})\right\rangle\right] \nonumber \\
    & = \sum_{t=1}^T\mathbb{E}_{\gamma}\left[\left\langle \left(\mathbf{s}_t-\frac{M}{N}\mathbb{I}_N\right),f(\mathbf{x}_t,\mathbf{s}_t)\right\rangle\right] + \underbrace{\sum_{t=1}^T\left[\phi_t(\mathbf{X}_t,\mathbf{Y}_{t+1})- \phi_t(\mathbf{X}_t,\mathbf{Y}_t)\right]}_{T_1} - \underbrace{\sum_{t = 1}^T\frac{1}{2}\left\langle \mathbf{y}_t, \nabla_\mathbf{Y}^2 \phi_t(\mathbf{X}_t, \Tilde{\mathbf{Y}}_t) \mathbf{y}_t\right\rangle}_{T_2},
    \label{eq:exppolicy}
\end{align}
In the above equation, the last equality comes from equation \eqref{eqn:Taylor}. Now, we can simplify the term $T_1$ as follows
\begin{align}
T_1:=\sum_{t=1}^T\left[\phi_t(\mathbf{X}_t,\mathbf{Y}_{t+1})- \phi_t(\mathbf{X}_t,\mathbf{Y}_t)\right] 
= \phi_T(\mathbf{X}_T,\mathbf{Y}_{T+1}) + \underbrace{\sum_{t=1}^{T-1}\left[\phi_{t}(\mathbf{X}_{t} ,\mathbf{Y}_{t+1}) -\phi_{t+1}(\mathbf{X}_{t+1},\mathbf{Y}_{t+1})\right] }_{T_3}-\phi_{1}(\mathbf{X}_{1},\mathbf{Y}_{1}). \label{eqn:s2}
\end{align}
Consider the term $\phi_T(\mathbf{X}_T,\mathbf{Y}_{T+1})$. We have
\begin{align}
   \phi_T(\mathbf{X}_T,Y_{T+1}) &=  \mathbb{E}_{\gamma}\left[\min_{\mathbf{s} \in \mathcal{S}}\left\langle\left(\mathbf{s} - \frac{M}{N}\mathbb{I}_N\right),\sum_{i=1}^{T-1}f(\mathbf{x}_i,\mathbf{s}) -\mathbf{Y}_{T+1} + \eta_T \gamma \right\rangle\right] \nonumber \\
   & \leq  \mathbb{E}_{\gamma}\left[\min_{\mathbf{s} \in \mathcal{S}}\left\langle\left(\mathbf{s} - \frac{M}{N}\mathbb{I}_N\right),\sum_{i=1}^{T}f(\mathbf{x}_i,\mathbf{s}) -\mathbf{Y}_{T+1} + \eta_T \gamma \right\rangle \underbrace{-\min_{\mathbf{s}\in \mathcal{S}} \left\langle\left(\mathbf{s} - \frac{M}{N}\mathbb{I}_N\right),f(\mathbf{x}_T,\mathbf{s})  \right\rangle}_{\leq 0}\right] \label{eq:32} \\
  &\leq \mathbb{E}_{\gamma}\left[\min_{\mathbf{s} \in \mathcal{S}}\left\langle\left(\mathbf{s} - \frac{M}{N}\mathbb{I}_N\right),\sum_{i=1}^{T}f(\mathbf{x}_i,\mathbf{s}) -\mathbf{Y}_{T+1} + \eta_T \gamma \right\rangle\right] \nonumber  \\
  & \leq \min_{\mathbf{s} \in \mathcal{S}}\mathbb{E}_{\gamma}\left[\left\langle\left(\mathbf{s} - \frac{M}{N}\mathbb{I}_N\right),\sum_{i=1}^{T}f(\mathbf{x}_i,\mathbf{s}) -\mathbf{Y}_{T+1} + \eta_T \gamma \right\rangle\right]  \label{eq:34} \\
  & = \min_{\mathbf{s} \in \mathcal{S}}\left\langle\left(\mathbf{s} - \frac{M}{N}\mathbb{I}_N\right),\sum_{i=1}^{T}f(\mathbf{x}_i,\mathbf{s}) -\mathbf{Y}_{T+1}  \right\rangle \\
  & =  \min_{\mathbf{s} \in \mathcal{S}}\left\langle\left(\mathbf{s} - \frac{M}{N}\mathbb{I}_N\right),\sum_{i=1}^{T}(f(\mathbf{x}_i,\mathbf{s}) -\mathbf{y}_{i})  \right\rangle
  \label{eq:35} 
\end{align}
Here, equation \eqref{eq:32} comes from $\min_x(f_1(x))+\min_x(f_2(x))\leq \min_x(f_1(x)+f_2(x))$ and the second term in equation \eqref{eq:32} is less than or equal to zero, since the term $\left\langle\left(\mathbf{s} - \frac{M}{N}\mathbb{I}_N\right),f(\mathbf{x}_T,\mathbf{s})  \right\rangle$  is the coded message size $R_{C}(\mathbf{s},x_T)$ and is greater than or equal to 0. Equation \eqref{eq:34} follows from Jensen's inequality and equation \eqref{eq:35} comes from the fact that $\mathbb{E}_{\gamma}[\gamma] = 0$. 

Now, consider the term $T_3$ in equation \eqref{eqn:s2}. We have
\begin{align}
    \phi_{t}(\mathbf{X}_{t} ,\mathbf{Y}_{t+1}) -\phi_{t+1}(\mathbf{X}_{t+1},\mathbf{Y}_{t+1}) &=  \mathbb{E}_{\gamma}\Bigg[\min_{\mathbf{s} \in \mathcal{S}}\left\langle\left(\mathbf{s} - \frac{M}{N}\mathbb{I}_N\right),\sum_{i=1}^{t-1}f(\mathbf{x}_i,\mathbf{s}) -\mathbf{Y}_{t+1} + \eta_t \gamma \right\rangle \nonumber \\
    & \hspace{1in}-\min_{\mathbf{s} \in \mathcal{S}}\left\langle\left(\mathbf{s} - \frac{M}{N}\mathbb{I}_N\right),\sum_{i=1}^{t}f(\mathbf{x}_i,\mathbf{s}) -\mathbf{Y}_{t+1} + \eta_{t+1} \gamma \right\rangle \Bigg] \nonumber \\
       & = \mathbb{E}_{\gamma}\Bigg[\min_{\mathbf{s} \in \mathcal{S}}\left\langle\left(\mathbf{s} - \frac{M}{N}\mathbb{I}_N\right),\sum_{i=1}^{t-1}f(\mathbf{x}_i,\mathbf{s}) -\mathbf{Y}_{t+1} + \eta_t \gamma \right\rangle \nonumber \\  
       &\hspace{1in} -\left\langle\left(\mathbf{s}_{t+1} - \frac{M}{N}\mathbb{I}_N\right),\sum_{i=1}^{t}f(\mathbf{x}_i,\mathbf{s}_{t+1}) -\mathbf{Y}_{t+1} + \eta_{t+1} \gamma \right\rangle\Bigg] \label{eq:st} \\
    & \leq -\mathbb{E}_{\gamma} \left[\min_{\mathbf{s}\in\mathcal{S}}\left\langle \left(\mathbf{s}-\frac{M}{N}\mathbb{I}_N\right),(\eta_{t+1}-\eta_{t})\gamma\right\rangle\right] \nonumber \\
    & \hspace{1in}- \mathbb{E}_{\gamma} \left[\left \langle\left(\mathbf{s}_{t+1}-\frac{M}{N}\mathbb{I}_N\right),f(\mathbf{x}_{t},\mathbf{s}_{t+1})\right\rangle\right] \label{eq:ff} \\
     & = -\mathbb{E}_{\gamma} \left[\min_{\mathbf{s}\in\mathcal{S}}\left\langle \mathbf{s}, (\eta_{t+1}-\eta_{t})\gamma\right\rangle\right] - \mathbb{E}_{\gamma} \left[\left\langle\left(\mathbf{s}_{t+1}-\frac{M}{N}\mathbb{I}_N\right),f(\mathbf{x}_{t},\mathbf{s}_{t+1})\right\rangle\right] \label{eqn:s1}\\
     &\leq |\eta_{t+1}-\eta_t|\mathcal{G}_{\max}(\gamma)- \mathbb{E}_{\gamma} \left[\left\langle\left(\mathbf{s}_{t+1}-\frac{M}{N}\mathbb{I}_N\right),f(\mathbf{x}_{t},\mathbf{s}_{t+1})\right\rangle\right]\label{eq:26}
\end{align}

Here,
\begin{itemize}
    \item Equation \eqref{eq:st} comes from the fact that, according to Algorithm \ref{alg:FTPL}, $$s_{t+1}=\arg\min_{\mathbf{s} \in \mathcal{S}}\left\langle\left(\mathbf{s} - \frac{M}{N}\mathbb{I}_N\right),\sum_{i=1}^{t}f(\mathbf{x}_i,\mathbf{s}) -\mathbf{Y}_{t+1} + \eta_{t+1} \gamma \right\rangle. $$
    \item Equation \eqref{eq:ff} comes from the following steps:
    
    \begin{align}
  - &\mathbb{E}_{\gamma} \left[\left\langle\left(\mathbf{s}_{t+1} - \frac{M}{N}\mathbb{I}_N\right),\sum_{i=1}^{t}f(\mathbf{x}_i,\mathbf{s}_{t+1}) -\mathbf{Y}_{t+1} + \eta_{t+1} \gamma \right\rangle\right]  \nonumber \\
  &\hspace{0.25in} = - \mathbb{E}_{\gamma} \left[\left\langle\left(\mathbf{s}_{t+1} - \frac{M}{N}\mathbb{I}_N\right),\sum_{i=1}^{t-1}f(\mathbf{x}_i,\mathbf{s}_{t+1}) -\mathbf{Y}_{t+1} + \eta_{t+1}  \gamma \right\rangle\right] - \mathbb{E}_{\gamma} \left[\left(\mathbf{s}_{t+1}-\frac{M}{N}\mathbb{I}_N\right)f(\mathbf{x}_{t},\mathbf{s}_{t+1})\right]\nonumber \\
  & \hspace{0.25in} \leq - \mathbb{E}_{\gamma} \left[\min_{\mathbf{s}\in \mathcal{S}}\left\langle\left(\mathbf{s} - \frac{M}{N}\mathbb{I}_N\right),\sum_{i=1}^{t-1}f(\mathbf{x}_i,\mathbf{s}) -\mathbf{Y}_{t+1} + \eta_{t+1}  \gamma \right\rangle\right] - \mathbb{E}_{\gamma} \left[\left(\mathbf{s}_{t+1}-\frac{M}{N}\mathbb{I}_N\right)f(\mathbf{x}_{t},\mathbf{s}_{t+1})\right] \nonumber \\
  &\hspace{0.25in} \leq - \mathbb{E}_{\gamma} \left[\min_{\mathbf{s}\in \mathcal{S}}\left\langle\left(\mathbf{s} - \frac{M}{N}\mathbb{I}_N\right),\sum_{i=1}^{t-1}f(\mathbf{x}_i,\mathbf{s}) -\mathbf{Y}_{t+1} + \eta_{t}  \gamma \right\rangle\right] - \mathbb{E}_{\gamma} \left[\left(\mathbf{s}_{t+1}-\frac{M}{N}\mathbb{I}_N\right)f(\mathbf{x}_{t},\mathbf{s}_{t+1})\right]- \nonumber \\ &\hspace{4in}\mathbb{E}_{\gamma} \left[\min_{\mathbf{s}\in \mathcal{S}}\left(\mathbf{s}-\frac{M}{N}\mathbb{I}_N\right)(\eta_{t+1}-\eta_{t}).\gamma\right]\label{eq:30}
\end{align}
    where, Equation \eqref{eq:30} follows from the fact $\min_x(f_1(x))+\min_x(f_2(x))\leq \min_x(f_1(x)+f_2(x))$.
\item Equation \eqref{eqn:s1} comes from the fact that $\mathbb{E}[\gamma] = 0$.
\end{itemize}
By substituting \eqref{eq:35} and \eqref{eq:26} in \eqref{eqn:s2}, we get
%
\begin{align}
     T_1 &= \phi_T(\mathbf{X}_T,Y_{T+1}) + \underbrace{\sum_{t=1}^{T-1}\left[\phi_{t}(\mathbf{X}_{t} ,Y_{t+1}) -\phi_{t+1}(\mathbf{X}_{t+1},Y_{t+1})\right] }_{T_3}-\phi_{1}(\mathbf{X}_{1},Y_{1}) \nonumber \\
    & \leq \phi_T(\mathbf{X}_T,Y_{T+1}) + \underbrace{\sum_{t=1}^{T-1} \left[  |\eta_{t+1}-\eta_t|\mathcal{G}_{\max}(\gamma) - \mathbb{E}_{\gamma} \left[\left(\mathbf{s}_{t+1}-\frac{M}{N}\mathbb{I}_N\right)f(\mathbf{x}_{t},\mathbf{s}_{t+1})\right] \right] }_{\geq T_3}-\phi_{1}(\mathbf{X}_{1},Y_{1}) \nonumber\\
       & \leq \min_{\mathbf{s} \in \mathcal{S}}\left\langle\left(\mathbf{s} - \frac{M}{N}\mathbb{I}_N\right),\sum_{i=1}^{T}f(\mathbf{x}_i,\mathbf{s}) -\mathbf{Y}_{T+1}  \right\rangle + (\eta_T-\eta_1)\mathcal{G}_{\max}(\gamma) - \nonumber \\
       &\hspace{2in}\sum_{t=1}^{T-1}\mathbb{E}_{\gamma} \left[\left(\mathbf{s}_{t+1}-\frac{M}{N}\mathbb{I}_N\right)f(\mathbf{x}_{t},\mathbf{s}_{t+1})\right] \underbrace{-\phi_{1}(\mathbf{X}_{1},Y_{1})}_{\eta_1\mathcal{G}_{\max}(\gamma)}
    \label{eq:38} 
\end{align}
{Note that $\mathbf{X}_1 = \mathbf{Y}_1 = \hat{0}$. Therefore, the term $-\phi_1(\mathbf{X}_1, \mathbf{Y}_1)$ in \eqref{eq:38} is $$-\phi_1(\mathbf{X}_1, \mathbf{Y}_1) = - \mathbb{E}_{\gamma}\left[\min\limits_{\mathbf{s} \in \mathcal{S}}\left\langle\left(\mathbf{s} - \frac{M}{N}\mathbb{I}_N\right), \eta_1 \gamma \right\rangle\right]  = - \mathbb{E}_{\gamma}\left[\min\limits_{\mathbf{s} \in \mathcal{S}}\left\langle\mathbf{s},\eta_1 \gamma \right\rangle\right] = \eta_1\mathcal{G}_{\max}(\gamma).$$}
Therefore,
\begin{align}
      T_1  \leq \min_{\mathbf{s} \in \mathcal{S}}\left\langle\left(\mathbf{s} - \frac{M}{N}\mathbb{I}_N\right),\sum_{i=1}^{T}f(\mathbf{x}_i,\mathbf{s}) -\mathbf{Y}_{T+1}  \right\rangle  + \eta_T\mathcal{G}_{\max}(\gamma)  - \sum_{t=1}^{T-1}\mathbb{E}_{\gamma} \left[\left(\mathbf{s}_{t+1}-\frac{M}{N}\mathbb{I}_N\right)f(\mathbf{x}_{t},\mathbf{s}_{t+1})\right] 
    \label{eq:38_2}
\end{align}
{For an $N$-dimensional vector $\mathbf{a}\in \mathbb{R}$ let $(\mathbf{a})_i\in \mathbb{R}$ denote its $i^{th}$ entry. Lastly, $T_2$ can be bounded as 
\begin{align}
    -\langle \mathbf{y}_t, \nabla^2 \phi_t(\mathbf{X}_t,\Tilde{\mathbf{Y}_t})\mathbf{y}_t\rangle \leq \left|\sum_{1=1}^N\sum_{j=1}^N (\mathbf{y}_t)_i(\mathbf{y}_t)_j|\nabla^2 \phi_t(\mathbf{X}_t,\Tilde{Y_t})|_{ij}\right|
\end{align}
As explained in Lemma 7 of \cite{abernethy2014online} and done in section 7.3 of \cite{bhattacharjee2020fundamental}
\begin{equation}
    \nabla^2 \phi_t(\mathbf{X}_t,\Tilde{\mathbf{Y}_t})_{ij} = \frac{1}{\eta_t}\mathbb{E}_{\gamma} [\nabla \widehat{\phi_t}(\mathbf{X}_t,\Tilde{\mathbf{Y}}_t-\eta\gamma)_i\gamma_j]
\end{equation}
where
\begin{equation}
    \widehat{\phi_t}(\mathbf{X},\mathbf{Y}) = \min_{\mathbf{s} \in \mathcal{S}}\left\langle\left(\mathbf{s} - \frac{M}{N}\mathbb{I}_N\right),\sum_{i=1}^{t-1}f(\mathbf{x}_i,\mathbf{s}) -\mathbf{Y}\right\rangle
\end{equation}
Thus $|\nabla^2 \phi_t(\mathbf{X}_t,\Tilde{\mathbf{Y}_t})|_{ij} \leq \frac{1}{\eta_t}\mathbb{E}_{\gamma} [|\nabla \widehat{\phi_t}(\mathbf{X}_t,\Tilde{\mathbf{Y}}_t-\eta\gamma)_i||\gamma_j|] \leq \frac{1}{\eta_t}\mathbb{E}_{\gamma} \left[\left|\left(\mathbf{s}(\mathbf{X}_t,\Tilde{\mathbf{Y}}_t-\eta_t\gamma)-\frac{M}{N}\mathbb{I}_N\right)_i\right||\gamma_j|\right] \leq \frac{\max\left\{\frac{M}{N},\left(1-\frac{M}{N}\right)\right\}}{\eta_t}\mathbb{E}_{\gamma}[|\gamma_j|]  = \frac{\max\left\{\frac{M}{N},\left(1-\frac{M}{N}\right)\right\}}{\eta_t}\sqrt{\frac{2}{\pi}}$. The last step comes from the fact that the max absolute value of an entry in $\left(\mathbf{s}(\mathbf{X}_t, \mathbf{Y}_t-\eta_t\gamma)-\frac{M}{N}\mathbb{I}_N\right)$ is $\left(1-\frac{M}{N}\right)$ for $2M\leq N$ and $\frac{M}{N}$ for $2M>N$ since $\mathbf{s}\in \mathcal{S}\subseteq \{0,1\}^N$. Here $\mathbf{s}(\mathbf{X}_t,\mathbf{Y}_t+\eta_t\gamma)_i$ is the $i^{th}$ entry of the vector with $\mathbf{s}$ chosen under inputs $\mathbf{X}_t$ and $Y_t - \eta\gamma$ to minimize $\widehat{\phi_t}(\mathbf{X},\mathbf{Y})$
\begin{align}
    -\langle \mathbf{y}_t, \nabla^2 \phi_t(\mathbf{x}_t,\Tilde{\mathbf{Y}_t})\mathbf{y}_t\rangle\leq \frac{\max\left\{\frac{M}{N},\left(1-\frac{M}{N}\right)\right\}}{\eta_t}\sqrt{\frac{2}{\pi}}\left|\sum_{1=1}^N\sum_{j=1}^N (\mathbf{y_t})_i(\mathbf{y_t})_j\right| \leq\frac{K^2\max\left\{\frac{M}{N},\left(1-\frac{M}{N}\right)\right\}}{\eta_t}\sqrt{\frac{2}{\pi}}
    \label{eq:41}
\end{align}
since
\begin{equation}
    \left|\sum_{1=1}^N\sum_{j=1}^N (\mathbf{y_t})_i(\mathbf{y_t})_j\right| \leq K^2 
\end{equation}
This follows because we have at most $K$ distinct requests at any time $t$. Thus 
\begin{equation}
    T_2\leq \frac{K^2\max\left\{\frac{M}{N},\left(1-\frac{M}{N}\right)\right\}}{\sqrt{2\pi}}\sum\limits_{t=1}^T\frac{1}{\eta_t}.
    \label{eq:t2bound}
\end{equation}
} From equations \eqref{eq:7}, \eqref{eq:exppolicy}, \eqref{eq:38_2} and \eqref{eq:t2bound}, we have  
\begin{align}
    R_{\text{UR}}((\mathbf{x}_t)_{t=1}^T,T)  &  \leq   
   \underbrace{\sum_{t=1}^{T-1}\mathbb{E}_{\gamma}\left[\left\langle \left(\mathbf{s}_t-\frac{M}{N}\mathbb{I}_N\right),f(\mathbf{x}_t,\mathbf{s}_t)\right\rangle - {\left\langle \left(\mathbf{s}_{t+1}-\frac{M}{N}\mathbb{I}_N\right), f(\mathbf{x}_{t},\mathbf{s}_{t+1})\right\rangle}\right]}_{T_5} + 
   \nonumber\\ 
   & \quad \quad \mathbb{E}_{\gamma}\left[\left\langle \left(\mathbf{s}_T-\frac{M}{N}\mathbb{I}_N\right),f(\mathbf{x}_T,\mathbf{s}_T)\right\rangle\right]+ \eta_T\mathcal{G}_{\max}(\gamma)+ \min_{\mathbf{s}\in\mathcal{S}}\left\langle\left(\mathbf{s} - \frac{M}{N}\mathbb{I}_N\right), \sum_{t=1}^{T}\left(f(\mathbf{x}_t,\mathbf{s}) - \mathbf{y}_t\right)\right\rangle \nonumber\\ 
   & \quad \quad +\frac{K^2\max\left\{\frac{M}{N},\left(1-\frac{M}{N}\right)\right\}}{\sqrt{2\pi}}\sum_{t=1}^T\frac{1}{\eta_t} + \eta_1\mathcal{G}_{\max}(\gamma) -    \min_{\mathbf{s}\in\mathcal{S}}\left\langle\left(\mathbf{s} - \frac{M}{N}\mathbb{I}_N\right), \sum_{t=1}^{T}\left(f(\mathbf{x}_t,\mathbf{s}) - \mathbf{y}_t\right)\right\rangle\\
     & \leq r_{\max}^C +\underbrace{\sum_{t=1}^{T-1}\mathbb{E}_{\gamma}\left[\left\langle \left(\mathbf{s}_t-\frac{M}{N}\mathbb{I}_N\right),f(\mathbf{x}_t,\mathbf{s}_t)\right\rangle - \left\langle \left(\mathbf{s}_{t+1}-\frac{M}{N}\mathbb{I}_N\right)f(\mathbf{x}_{t},\mathbf{s}_{t+1})\right\rangle\right]}_{T_5}  \nonumber \\
     & \hspace{2in} +\eta_T\mathcal{G}_{\max}(\gamma)+ \frac{K^2\max\left\{\frac{M}{N},\left(1-\frac{M}{N}\right)\right\}}{\sqrt{2\pi}}\sum_{t=1}^T\frac{1}{\eta_t} + \eta_1\mathcal{G}_{\max}(\gamma) \label{eqn:s6}
\end{align}

Note that we bound $\mathbb{E}_{\gamma}\left[\left\langle \left(\mathbf{s}_T-\frac{M}{N}\mathbb{I}_N\right),f(\mathbf{x}_T,\mathbf{s}_T)\right\rangle\right]$ by $r_{\max}^C$ to obtain equation \eqref{eqn:s6}.

Now consider the term $T_5$. We have
\begin{align}
    T_5  &:= \sum_{t=1}^{T-1}\mathbb{E}_{\gamma}\left[\left\langle \left(\mathbf{s}_t-\frac{M}{N}\mathbb{I}_N\right),f(\mathbf{x}_t,\mathbf{s}_t)\right\rangle - \left\langle \left(\mathbf{s}_{t+1}-\frac{M}{N}\mathbb{I}_N\right)f(\mathbf{x}_{t},\mathbf{s}_{t+1})\right\rangle\right] \nonumber \\
     &= \sum_{t=1}^{T-1}\mathbb{E}_{\gamma}\left[ \left( \left\langle\left(\mathbf{s}_t-\frac{M}{N}\mathbb{I}_N\right),f(\mathbf{x}_t,\mathbf{s}_t)\right\rangle - \left\langle \left(\mathbf{s}_{t+1}-\frac{M}{N}\mathbb{I}_N\right)f(\mathbf{x}_{t},\mathbf{s}_{t+1})\right\rangle\right)\mathbb{I}(\mathbf{s}_{t+1}\neq \mathbf{s}_t)\right] \label{eqn:s3}\\
     & \leq \sum_{t=1}^{T-1}\mathbb{E}_{\gamma}\left[r_{\max}^C\mathbb{I}(\mathbf{s}_{t+1}\neq \mathbf{s}_t)\right] \label{eqn:s4}\\
     & = r_{\max}^C \sum_{t=1}^{T-1}\mathbb{E}_{\gamma}\left[\mathbb{I}(\mathbf{s}_{t+1}\neq \mathbf{s}_t)\right] \label{eq:preswitchlemma} \\
     &= \frac{3r_{\max}r_{\max}^C(|\mathcal{S}|-1)}{2\sqrt{2\pi}\alpha}\sum_{t=1}^T\frac{1}{\sqrt{t}} \label{eqn:s5}
\end{align}
Here, 
\begin{itemize}
    \item Equation \eqref{eqn:s3} comes from the fact that the terms inside the $T_5$ summation are zero when $\mathbf{s}_t = \mathbf{s}_{t+1}$.
    \item Equation \eqref{eqn:s4} comes from the fact that {the term $\left\langle \left(\mathbf{s}_t-\frac{M}{N}\mathbb{I}_N\right),f(\mathbf{x}_t,\mathbf{s}_t)\right\rangle - \left\langle \left(\mathbf{s}_{t+1}-\frac{M}{N}\mathbb{I}_N\right)f(\mathbf{x}_{t},\mathbf{s}_{t+1})\right\rangle$ is essentially the difference in expected coded rates $R_{C}(\mathbf{s}_t,\mathbf{x}_t)$ and $R_{C}(\mathbf{s}_{t+1},\mathbf{x}_t)$ which can be upper bounded by max coded rate $r_{\max}^C$.}
    \item Equation \eqref{eqn:s5} comes from {lemma \ref{lemma:lemmaswitchff}.}
\end{itemize}
{
By substituting \eqref{eqn:s5} in \eqref{eqn:s6}, we get 
\begin{equation}
  R_{\text{UR}}((\mathbf{x}_t)_{t=1}^T,T)  \leq \frac{3r_{\max}r_{\max}^C(|\mathcal{S}|-1)}{2\sqrt{2\pi}\alpha}\sum_{t=1}^T\frac{1}{\sqrt{t}} + \eta_T\mathcal{G}_{\max}(\gamma)+\frac{K^2\max\left\{\frac{M}{N},\left(1-\frac{M}{N}\right)\right\}}{\sqrt{2\pi}}\sum_{t=1}^T\frac{1}{\eta_t} +  \eta_1\mathcal{G}_{\max}(\gamma)+ r_{\max}^C\label{eq:45}
\end{equation}
Since this is true for every sequence of request patterns $(\mathbf{x}_t)_{t=1}^T$ we have
\begin{equation}
  R_{\text{UR}}(T)  \leq \frac{3r_{\max}r_{\max}^C(|\mathcal{S}|-1)}{2\sqrt{2\pi}\alpha}\sum_{t=1}^T\frac{1}{\sqrt{t}} + \eta_T\mathcal{G}_{\max}(\gamma)+\frac{K^2\max\left\{\frac{M}{N},\left(1-\frac{M}{N}\right)\right\}}{\sqrt{2\pi}}\sum_{t=1}^T\frac{1}{\eta_t} +  \eta_1\mathcal{G}_{\max}(\gamma)+ r_{\max}^C\label{eq:45_2}
\end{equation}
Theorem \ref{theorem:regret_unrestricted} follows by replacing $\max\left\{\frac{M}{N},\left(1-\frac{M}{N}\right)\right\}\leq 1$, $r_{\max}\leq K$ and $r_{\max}^C\leq K$ with their respective upper bounds.}

\end{proof}
\begin{corollary}
    The Regret upper bound in Theorem \ref{theorem:regret_unrestricted} for a general non-decreasing learning rate schedule $\{\eta_t\}_{\{t = 1:T\}}$ can be given by 
    \begin{equation}
          R_{\text{UR}}(T)  \leq \frac{\sqrt{2}r_{\max}r_{\max}^C(|\mathcal{S}|-1)}{\sqrt{\pi}\alpha}\sum_{t=1}^T\frac{1}{\eta_t} + \eta_T\mathcal{G}_{\max}(\gamma)+\frac{K^2\max\left\{\frac{M}{N},\left(1-\frac{M}{N}\right)\right\}}{\sqrt{2\pi}}\sum_{t=1}^T\frac{1}{\eta_t} +  \eta_1\mathcal{G}_{\max}(\gamma)+ r_{\max}^C
    \end{equation}
\end{corollary}
\begin{proof}
Continuing from equation \eqref{eq:preswitchlemma} 
\begin{align}
    & = r_{\max}^C \sum_{t=1}^{T-1}\mathbb{E}_{\gamma}\left[\mathbb{I}(\mathbf{s}_{t+1}\neq \mathbf{s}_t)\right] \label{eq:preswitchlemma_gen} \\
     &= \frac{2r_{\max}r_{\max}^C(|\mathcal{S}|-1)}{\sqrt{2\pi}}\sum_{t=1}^T\frac{1}{\eta_t} \label{eqn:s5_gen}
\end{align}
 Here we used the upper bound from Corollary \ref{cor:switch} By substituting \eqref{eqn:s5_gen} in \eqref{eqn:s6}, we get 
\begin{equation}
  R_{\text{UR}}((\mathbf{x}_t)_{t=1}^T,T)  \leq \frac{\sqrt{2}r_{\max}r_{\max}^C(|\mathcal{S}|-1)}{\sqrt{\pi}\alpha}\sum_{t=1}^T\frac{1}{\eta_t} + \eta_T\mathcal{G}_{\max}(\gamma)+\frac{K^2\max\left\{\frac{M}{N},\left(1-\frac{M}{N}\right)\right\}}{\sqrt{2\pi}}\sum_{t=1}^T\frac{1}{\eta_t} +  \eta_1\mathcal{G}_{\max}(\gamma)+ r_{\max}^C\label{eq:45_gen}
\end{equation}
Since this is true for every sequence of request patterns $(\mathbf{x}_t)_{t=1}^T$ we have
\begin{equation}
  R_{\text{UR}}(T)  \leq \frac{\sqrt{2}r_{\max}r_{\max}^C(|\mathcal{S}|-1)}{\sqrt{\pi}\alpha}\sum_{t=1}^T\frac{1}{\eta_t} + \eta_T\mathcal{G}_{\max}(\gamma)+\frac{K^2\max\left\{\frac{M}{N},\left(1-\frac{M}{N}\right)\right\}}{\sqrt{2\pi}}\sum_{t=1}^T\frac{1}{\eta_t} +  \eta_1\mathcal{G}_{\max}(\gamma)+ r_{\max}^C\label{eq:45_2_gen}
\end{equation}
\end{proof}

\subsection{Proof of Theorem \ref{theorem:advswitch}}
\label{app: Th2switch}
We will use the following lemma to prove Theorem \ref{theorem:advswitch}.


\begin{lemma}\label{lemma:lemmaswitch}
For a given history of request vectors $(\mathbf{x}_i)_{i=1}^{t-1}$ up to time $t-1$ and a perturbation vector $\gamma$,
    the probability of switching cache configuration for Algorithm \ref{alg:FTPL} in the step $t$    for $\mathbf{s} \neq \mathbf{s}_t$ is given by
{\begin{equation}
    \mathbb{P}\left((\mathbf{s}_{t+1} = \mathbf{s})|(\mathbf{x}_1\cdots \mathbf{x}_{t-1},\gamma)\right) \leq \frac{3r_{\max}}{2\alpha\sqrt{2\pi}\sqrt{t+1}}    
\end{equation}}
\end{lemma}

\begin{proof}   
Let $\mathbf{X}_t=(\mathbf{x}_i)_{i=1}^{t-1}$ and $\mathbf{Y}_{t}=\sum_{i=1}^{t-1}\min\{\mathbf{x}_{i}, \mathbb{I}_N\}$. For a given $\mathbf{X}_t$ and $\mathbf{Y}_t$, let $R_t(\mathbf{s},\mathbf{X}_t,\mathbf{Y}_t)$ denote the {cumulative transmission rate up to time $t$ under} a cache configuration $\mathbf{s}$ defined as 
\begin{equation}
    R_t(\mathbf{s},\mathbf{X}_t,\mathbf{Y}_t) = \left\langle\left(\mathbf{s} - \frac{M}{N}\mathbb{I}_N\right),\sum_{i=1}^{t-1}f(\mathbf{x}_i,\mathbf{s}) -\mathbf{Y}_t  \right\rangle.
\end{equation}
Let $\mathbf{s}_t$ and $\mathbf{s}_{t+1}$ denote the cache configurations of Algorithm \ref{alg:FTPL} in time slots $t$ and $t+1$ respectively. Consider a cache configuration $\mathbf{s}\neq \mathbf{s}_t$. If $\mathbf{s}_{t+1}=\mathbf{s}$, then we have the event that \remove{Note that the dimension of $\mathbf{X}$ input to $R_t$ is $N\times (t-1)$ and that of $\mathbf{s}$ and $\mathbf{Y}$ input is $N$. The function maps them to a scalar. For $\mathbf{s}\neq \mathbf{s}_{t}$, given $\left(\mathbf{x}_1\cdots \mathbf{x}_{t-1},\gamma)\right)$  at time $t$, following algorithm \ref{alg:FTPL} a switch can happen when for some $\mathbf{s}\in \mathcal{S}\setminus \mathbf{s}_t$ we have\\}
  $$\left\{R_t(\mathbf{s}_t,\mathbf{X}_t,\mathbf{Y}_t-\eta_t\gamma)\leq R_t(\mathbf{s},\mathbf{X}_t,\mathbf{Y}_t-\eta_t \gamma)\right\}\cap  \left\{R_{t+1}(\mathbf{s}_t,\mathbf{X}_{t+1},\mathbf{Y}_{t+1}-\eta_{t+1}\gamma)\geq R_{t+1}(\mathbf{s},\mathbf{X}_{t+1},\mathbf{Y}_{t+1}-\eta_{t+1}\gamma)\right\}.$$ 
  
  Let $\mathbb{P}_\gamma^{\mathcal{H}_t}(.)$ denote the function $\mathbb{P}_\gamma\left((.|(\mathbf{x}_1\cdots \mathbf{x}_{t-1},\gamma)\right)$.  Then,  we have 
  
\begin{align}
      \mathbb{P}_\gamma^{\mathcal{H}_t}(\mathbf{s}_{t+1} = \mathbf{s})  &\leq \mathbb{P}_\gamma^{\mathcal{H}_t}(\left\{R_t(\mathbf{s}_t,\mathbf{X}_t,\mathbf{Y}_t-\eta_t\gamma)\leq R_t(\mathbf{s},\mathbf{X}_t,\mathbf{Y}_t-\eta_t \gamma)\right\} \cap \nonumber \\
      &\hspace{2in}\left\{R_{t+1}(\mathbf{s}_t,\mathbf{X}_{t+1},\mathbf{Y}_{t+1}-\eta_{t+1}\gamma)\geq R_{t+1}(\mathbf{s},\mathbf{X}_{t+1},\mathbf{Y}_{t+1}-\eta_{t+1}\gamma)\right\}) \nonumber \\
   & = \mathbb{P}_\gamma^{\mathcal{H}_t}\bigg(\left\{R_t(\mathbf{s}_t,\mathbf{X}_t,\mathbf{Y}_t) + \left\langle\left(\mathbf{s}_t-\frac{M}{N}\mathbb{I}_N\right),\eta_t\gamma\right\rangle\leq R_t(\mathbf{s},\mathbf{X}_t,\mathbf{Y}_t)+\left\langle\left(\mathbf{s}-\frac{M}{N}\mathbb{I}_N\right),\eta_t\gamma\right\rangle \right\} \cap \nonumber \\ 
   &\quad \left\{R_{t+1}(\mathbf{s}_t,\mathbf{X}_{t+1},Y_{t+1}) + \left\langle\left(\mathbf{s}_t-\frac{M}{N}\mathbb{I}_N\right),\eta_{t+1}\gamma\right\rangle\geq R_{t+1}(\mathbf{s},\mathbf{X}_{t+1},\mathbf{Y}_{t+1})+ \left\langle\left(\mathbf{s}-\frac{M}{N}\mathbb{I}_N\right),\eta_{t+1}\gamma\right\rangle\right\}\bigg)\nonumber \\
    &= \mathbb{P}_\gamma^{\mathcal{H}_t}\left(\frac{R_{t}(\mathbf{s}_t,\mathbf{X}_{t},\mathbf{Y}_{t}) - R_{t}(\mathbf{s},\mathbf{X}_{t},\mathbf{Y}_{t})}{\eta_{t}} \leq \langle(\mathbf{s}-\mathbf{s}_t),\gamma\rangle 
    \leq \frac{R_{t+1}(\mathbf{s}_t,\mathbf{X}_{t+1},\mathbf{Y}_{t+1}) - R_{t+1}(\mathbf{s},\mathbf{X}_{t+1},\mathbf{Y}_{t+1})}{\eta_{t+1}}\right) \nonumber \\   
   & \leq  \mathbb{P}_\gamma^{\mathcal{H}_t}\left(\frac{R_{t}(\mathbf{s}_t,\mathbf{X}_{t},\mathbf{Y}_{t}) - R_{t}(\mathbf{s},\mathbf{X}_{t},\mathbf{Y}_{t})}{\eta_{t}} \leq \langle(\mathbf{s}-\mathbf{s}_t),\gamma\rangle \leq \frac{R_{t}(\mathbf{s}_t,\mathbf{X}_{t},\mathbf{Y}_{t}) +r_{\max} - R_{t}(\mathbf{s},\mathbf{X}_{t},\mathbf{Y}_{t})}{\eta_{t+1}}\right)\label{eq:50}\\
    & \leq  \frac{1}{\sqrt{2\pi}\sigma_{(\mathbf{s},\mathbf{s}_t)}}\left(\frac{r_{\max}}{\eta_{t+1}} + \left(\frac{1}{\eta_{t}} -\frac{1}{\eta_{t+1}} \right)(R_{t}(\mathbf{s}_{t+1},\mathbf{X}_{t},Y_{t}) - R_{t}(\mathbf{s}_{t},\mathbf{X}_{t},Y_{t}))\right)\label{eq:49}\\
    &\leq  \frac{1}{\sqrt{2\pi}\alpha}\left(\frac{r_{\max}}{\sqrt{t+1}} + \left(\frac{1}{\sqrt{t}} -\frac{1}{\sqrt{t+1}} \right)tr_{\max}\right)=\frac{1}{\sqrt{2\pi}\alpha}\left(\frac{r_{\max}}{\sqrt{t+1}} + \left( \frac{\sqrt{t}}{\sqrt{t+1}(\sqrt{t+1}+\sqrt{t})}\right)r_{\max}\right)\label{eq:51}\\
    &\leq  \frac{3r_{\max}}{2\alpha\sqrt{2\pi}\sqrt{t+1}} 
\end{align}

 {Equation \eqref{eq:50} can be obtained by expressing $R_t(\mathbf{s},\mathbf{X},\mathbf{Y})$ terms as a sum of $K(\mathbf{s},\mathbf{x})$ terms. $R_{t+1}(\mathbf{s}_t,\mathbf{x}_{t+1},\mathbf{Y}_{t+1}) - R_{t+1}(\mathbf{s},\mathbf{x}_{t+1},\mathbf{Y}_{t+1})$ = $(\sum_{i=1}^t K(\mathbf{s}_t,\mathbf{x}_t) - \sum_{i=1}^t h(\mathbf{x}_i))  - (\sum_{i=1}^t K(\mathbf{s},\mathbf{x}_t)-\sum_{i=1}^t h(\mathbf{x}_i))$. where $h(\mathbf{x}_t) = \left(1- \frac{M}{N}\right)\langle \mathbf{y}_t, \mathbb{I}_N\rangle$. The right side of the inequality in equation \ref{eq:50} follows as $\sum_{i=1}^t K(\mathbf{s}_t,\mathbf{x}_t) \leq \sum_{i=1}^{t-1} K(\mathbf{s}_t,\mathbf{x}_t)+r_{\max}$ and $\sum_{i=1}^t K(\mathbf{s},\mathbf{x}_t)\geq\sum_{i=1}^{t-1} K(\mathbf{s},\mathbf{x}_t))$ as the rate at time $t$ is positive.

Equation \eqref{eq:49} comes from the fact that the Gaussian PDF is upper bounded by $\frac{1}{\sqrt{2\pi}\sigma}$. Equation \ref{eq:51} comes from $\left\langle(\mathbf{s}-\mathbf{s}_t),\gamma\right\rangle$ being a sum of unit variance independent Gaussian random variables. Let $\sigma_{(s,\mathbf{s}_t)}$ be the standard deviation for a given $(\mathbf{s}_t,\mathbf{s})$ pair. The last step comes from choosing $\eta_t = \alpha\sqrt{t}$, and $\min\limits_{\mathbf{s}\neq \mathbf{s}_t}\sigma_{(\mathbf{s},\mathbf{s}_t)} = 1$ which happens when $(\mathbf{s}_t,\mathbf{s})$ differ at exactly one file. } 
\end{proof}
\begin{corollary}\label{cor:preswitch}
    For a general non-decreasing learning rate schedule $\{\eta_t\}$ lemma \ref{lemma:lemmaswitch} can be rewritten as
        \begin{equation}
    \mathbb{P}\left((\mathbf{s}_{t+1} = \mathbf{s})|(\mathbf{x}_1\cdots \mathbf{x}_{t-1},\gamma)\right) \leq \frac{1}{\sqrt{2\pi}}\left(\frac{r_{\max}}{\eta_{t+1}} + \left(\frac{1}{\eta_{t}} -\frac{1}{\eta_{t+1}} \right)(tr_{\max})\right)  
    \end{equation}
\end{corollary}
\begin{proof}
{Continuing from equation \eqref{eq:49}
\begin{align}
    \leq &  \frac{1}{\sqrt{2\pi}\sigma_{(\mathbf{s},\mathbf{s}_t)}}\left(\frac{r_{\max}}{\eta_{t+1}} + \left(\frac{1}{\eta_{t}} -\frac{1}{\eta_{t+1}} \right)(R_{t}(\mathbf{s}_{t+1},\mathbf{X}_{t},Y_{t}) - R_{t}(\mathbf{s}_{t},\mathbf{X}_{t},Y_{t}))\right)\label{eq:49_gen}\\
    \leq &  \frac{1}{\sqrt{2\pi}\sigma_{(\mathbf{s},\mathbf{s}_t)}}\left(\frac{r_{\max}}{\eta_{t+1}} + \left(\frac{1}{\eta_{t}} -\frac{1}{\eta_{t+1}} \right)(tr_{\max})\right)\label{eq:gen2}\\
    \leq &  \frac{1}{\sqrt{2\pi}}\left(\frac{r_{\max}}{\eta_{t+1}} + \left(\frac{1}{\eta_{t}} -\frac{1}{\eta_{t+1}} \right)(tr_{\max})\right)\label{eq:gen3}
\end{align}

}
\end{proof}

\begin{lemma}
\label{lemma:lemmaswitchff}
    For a coded caching problem with $N$ files, $K$ users/caches, and cache size $MF$ bits, let $C_{\text{UR} }(T)$ denote
the expected number of switches in cache configuration until time $T$ for Algorithm \ref{alg:FTPL} with unrestricted switching. Then
\begin{equation}
   C_{\text{UR} }(T) \leq \frac{3r_{\max}(|\mathcal{S}|-1)}{2\sqrt{2\pi}\alpha} \sum_{t=1}^T \frac{1}{\sqrt{t}}=\mathcal{O}(\sqrt{T}). \nonumber 
\end{equation}
\end{lemma}
\remove{\textbf{Theorem 2}\\
\label{thm:fullswitchadv}
The expected number of switches in cache configuration until time $T$ for Algorithm \ref{alg:FTPL} with unrestricted switching and input parameters $\mathbf{\eta}=(\eta_t=\alpha\sqrt{t})_{t\in \mathcal{T}}$ is given by
\begin{equation}
     C_{(\mathbf{\eta}, \text{UR}) }^A(T) \leq \frac{3r_{\max}(|\mathcal{S}|-1)}{2\sqrt{\pi}\alpha} \sum_{t=1}^T \frac{1}{\sqrt{t}} = \mathcal{O}(\sqrt{T}) \nonumber 
\end{equation}
where $|\mathcal{S}|$ is the cardinality of the set of feasible cache configurations.}
\begin{proof}
Recall that $C_{\text{UR} }(T)$ denote
the expected number of switches in cache configuration until time $T$ for Algorithm \ref{alg:FTPL} with unrestricted switching. Then, we have
\begin{align}
  C_{ \text{UR} }(T)   & = \mathbb{E}\left[\sum_{t=1}^{T-1}\mathbb{I}(\mathbf{s}_{t+1} \neq \mathbf{s}_t)\right]\\
    & =  \sum_{t=1}^{T-1}\mathbb{E}\left[\mathbb{E}\left[\mathbb{I}(\mathbf{s}_{t+1} \neq \mathbf{s}_t)|(\mathbf{x}_1\cdots \mathbf{x}_{t-1},\gamma)\right]\right]\\
    & =  \sum_{t=1}^{T-1}\mathbb{E}\left[\mathbb{P}\left((\mathbf{s}_{t+1} \neq \mathbf{s}_t)|(\mathbf{x}_1\cdots \mathbf{x}_{t-1},\gamma)\right)\right]\\
    & = \sum_{t=1}^{T-1}\sum_{\mathbf{s}\in \mathcal{S}\setminus \mathbf{s}_t}\mathbb{E}\left[\mathbb{P}\left((\mathbf{s}_{t+1} = \mathbf{s})|(\mathbf{x}_1\cdots \mathbf{x}_{t-1},\gamma)\right)\right]\label{eq:st_deterministic}\\ 
    & \leq \sum_{t=1}^{T-1}\sum_{\mathbf{s}\in \mathcal{S}\setminus \mathbf{s}_t} \mathbb{E}\left[\frac{3r_{\max}}{2\alpha\sqrt{2\pi}\sqrt{t+1}}\right]\label{eq: lemmaswitch}\\ 
    & \leq \frac{3r_{\max}(|\mathcal{S}|-1)}{2\sqrt{2\pi}\alpha} \sum_{t=1}^T\frac{1}{\sqrt{t}}    
\end{align}
{Here, equation \eqref{eq:st_deterministic} comes from the fact that for the policy (Algorithm \ref{alg:FTPL}) $\mathbf{s}_t$ can be determined given past observations $\mathbf{x}_t$ and the perturbation vector $\gamma$. Equation \eqref{eq: lemmaswitch} comes from the lemma \ref{lemma:lemmaswitch}. 
The proof of Theorem \ref{theorem:advswitch} can be obtained by replacing $r_{\max}$ by $K$ ($r_{\max}\leq K$) in lemma \ref{lemma:lemmaswitchff}}
\begin{corollary}\label{cor:switch}
    The upper bound on the switching cost for a general non-decreasing learning rate schedule $\{\eta_t\}$ is given by
\begin{equation}
   C_{\text{UR} }(T) \leq \frac{2r_{\max}(|\mathcal{S}|-1)}{\sqrt{2\pi}}\sum_{t=1}^{T}\frac{1}{\eta_{t}}. \nonumber 
\end{equation}
\end{corollary}
{
Continuing from equation \eqref{eq:st_deterministic} using the expression from \eqref{eq:gen3}
\begin{align}
=& \sum_{t=1}^{T-1}\sum_{\mathbf{s}\in \mathcal{S}\setminus \mathbf{s}_t}\mathbb{E}\left[\mathbb{P}\left((\mathbf{s}_{t+1} = \mathbf{s})|(\mathbf{x}_1\cdots \mathbf{x}_{t-1},\gamma)\right)\right]\label{eq:st_deterministic_gen}\\
\leq & \sum_{t=1}^{T-1}\sum_{\mathbf{s}\in \mathcal{S}\setminus \mathbf{s}_t} \mathbb{E}\left[\frac{1}{\sqrt{2\pi}}\left(\frac{r_{\max}}{\eta_{t+1}} + \left(\frac{1}{\eta_{t}} -\frac{1}{\eta_{t+1}} \right)(tr_{\max})\right)\right]\label{eq: lemmaswitch_gen}\\
= & \frac{r_{\max}}{\sqrt{2\pi}}\left(\sum_{\mathbf{s}\in \mathcal{S}\setminus \mathbf{s}_t}\sum_{t=1}^{T-1}\frac{1}{\eta_{t+1}} + \sum_{t=1}^{T-1}\left(\frac{1}{\eta_{t}} -\frac{1}{\eta_{t+1}} \right)(t)\right)\\
\leq & \frac{r_{\max}(|\mathcal{S}|-1)}{\sqrt{2\pi}}\left(\sum_{t=2}^{T}\frac{1}{\eta_{t}} + \sum_{t=1}^{T-1}\frac{1}{\eta_t}\right)\\
\leq & \frac{2r_{\max}(|\mathcal{S}|-1)}{\sqrt{2\pi}}\sum_{t=1}^{T}\frac{1}{\eta_{t}} 
\end{align}
\label{eq:switchgen}
}
In order to obtain equation \eqref{eq: lemmaswitch_gen} we use the upper bound from Corollary \ref{cor:preswitch}
\end{proof}
\subsection{Proof of Theorem \ref{theorem:regret_restricted_switching}}
\label{app: resadv}
\remove{In this scenario, the cache updates are allowed only at certain time steps. These time steps are indicated by inter-switch periods $l_i$ and $\sum_{i=1}^L l_i = T$. The $k^{th}$ cache update is allowed during the $\sum_{i=1}^{k-1}l_i$.\\
\textbf{Theorem 3:}\\
For an adversarial coded caching problem with $N$ files, $K$ users/caches, and cache size $MF$ bits, let $R_{(\eta, \text{R})}^{\mathcal{T}}(T)$ be the adversarial regret of Algorithm \ref{alg:FTPL} under restricted switching with switching slots $\mathcal{T}$ and input parameters $\mathbf{\eta}=(\eta_t=\alpha\sqrt{t})_{t\in \mathcal{T}}$. Then,
\begin{equation}
R_{(\eta, \text{R})}^{\mathcal{T}}(T) \leq R_{(\eta, \text{UR})}(T) +  \sum_{k=1}^{L}\frac{3r_{\max}^2(|\mathcal{S}|-1)l_k(l_k-1)}{4\alpha\sqrt{\pi}\sqrt{\sum_{i=1}^{k-1} l_i+1}}. \nonumber
\end{equation}}

\begin{proof}
    Theorem \ref{theorem:advswitch} (proof) suggests that cache configurations do not change frequently. Note that Algorithm \ref{alg:FTPL} is the same for both the unrestricted switching scenario and the restricted switching scenario. However, in the unrestricted switching scenario, the cache configuration is allowed to change in every time slot $t\in [T]$, whereas in the restricted switching scenario, the cache configuration is allowed to change only in restricted time slots given by the set $\mathcal{T}$. Therefore, the difference in the transmission sizes incurred in these two cases solely comes from the cache switches that take place in between these inter-switch periods.
 
    Recall that $l_k \triangleq t_k-t_{k-1}$ denotes the time gap between the $(k-1)^{th}$ and $k^{th}$ switching slots, and $L \triangleq |\mathcal{T}|$ denotes the maximum number of allowed switches.
For the restricted switching scenario, we enjoy the same transmission size starting from $\sum_{i=1}^{k-1}l_i$ as the transmission size for the unrestricted scenario until there is a switch at some time $t$ for the unrestricted scenario, where $\sum_{i=1}^{k-1} l_i < t < \sum_{i=1}^{k}l_{i}$. Once a switch happens at instant $\sum_{i=1}^{k-1} l_i < t < \sum_{i=1}^{k}l_{i}$, there is no guarantee that the regret incurred for the restricted switching scenario from time $t$ to  $\sum_{i=1}^{k}l_i -1$ is the same as that of the unrestricted switching scenario.

    Let $R_k$ be the total additional regret (compared to the unrestricted switching scenario) incurred during the $k^{th}$ inter-switch period. Then, $R_{(\eta, \text{R})}^{\mathcal{T}}(T) = R_{(\eta, \text{UR})}^{\mathcal{T}}(T) + \sum_{k=1}^L R_k$. If the switch happens for the first time after $\sum_{i=1}^{k-1} l_i$ at time-slot $t$, we have the regret in $R_k$ being upper bounded by $\left(\sum_{i=1}^k l_i -t\right)r_{\max}$. Therefore, we have

    \begin{align}
        R_k &\leq \sum_{t=\sum_{i=1}^{k-1} l_i+1}^{\sum_{i=1}^k l_i-1 }\mathbb{E}\left[\mathbb{I}(\mathbf{s}_{t}\neq \mathbf{s}_{t-1})\right]\left(\sum_{i=1}^k l_i -t\right)r_{\max}\nonumber \\
         & = \sum_{t=\sum_{i=1}^{k-1} l_i+1}^{\sum_{i=1}^k l_i-1 }\mathbb{E}\left[\mathbb{E}\left[\mathbb{I}(\mathbf{s}_{t}\neq \mathbf{s}_{t-1})| (\mathbf{x}_1 \cdots \mathbf{x}_{t-1}, \gamma)\right]\right]\left(\sum_{i=1}^k l_i -t\right)r_{\max} \nonumber \\
        & = \sum_{t=\sum_{i=1}^{k-1} l_i+1}^{\sum_{i=1}^k l_i-1 }\mathbb{E}\left[\sum_{\mathbf{s}\in \mathcal{S}\setminus \mathbf{s}_{t-1}}\mathbb{P}\left((\mathbf{s}_{t} = \mathbf{s})|(\mathbf{x}_1\cdots \mathbf{x}_{t-2},\gamma)\right)\right]\left(\sum_{i=1}^k l_i -t\right)r_{\max}\label{eq: lemmares}\\
        & \leq \sum_{t=\sum_{i=1}^{k-1} l_i+1}^{\sum_{i=1}^k l_i-1 }\sum_{\mathbf{s}\in \mathcal{S}\setminus \mathbf{s}_{t-1}}\frac{3r_{\max}}{2\alpha\sqrt{\pi}\sqrt{t}}\left(\sum_{i=1}^k l_i -t\right)r_{\max}\\
        & \leq \frac{3r_{\max}^2(|\mathcal{S}|-1)}{2\alpha\sqrt{\pi}\sqrt{\sum_{i=1}^{k-1} l_i+1}}\sum_{t=\sum_{i=1}^{k-1} l_i+1}^{\sum_{i=1}^k l_i-1 }\left(\sum_{i=1}^k l_i - t\right)\label{eq:lileq}\\
        & {=} \frac{3r_{\max}^2(|\mathcal{S}|-1)}{2\alpha\sqrt{\pi}\sqrt{\sum_{i=1}^{k-1} l_i+1}}\frac{l_k(l_k-1)}{2}
    \end{align}
{Equation \eqref{eq: lemmares} come from lemma \ref{lemma:lemmaswitch} and equation \eqref{eq:lileq} comes from $\sum\limits_{i=1}^{k-1} l_i <t$ .Adding all these $R_k$ terms gives the above result.}
\end{proof}
\subsection{Linear approximation}
\label{app:linapp}
One of the benchmarks we compare against is a linear approximation of our scheme. This is the same approximation as the one used in \cite{nayak2023regret}. Recall that as per Proposition \ref{proposition:expected_rate} the expected rate incurred by the cache configuration $\mathbf{s}_t$ for a request vector $\mathbf{x}_t$ can be given by
\begin{equation}
        K(\mathbf{s}_t,\mathbf{x}_t) =\underbrace{\langle(\mathbb{I}_N-\mathbf{s}_t),\mathbf{y}_t\rangle}_{\text{Uncoded transmission}} +\underbrace{\left(\frac{\langle \mathbf{s}_t, \mathbb{I}_N\rangle}{M}-1\right)\left(1-\left(1-\frac{M}{\langle \mathbf{s}_t, \mathbb{I}_N\rangle}\right)^{\langle \mathbf{x}_t,\mathbf{s}_t\rangle}\right)}_{\text{Coded transmission}}
\end{equation}
For the regime where $\langle \mathbf{x}_t,\mathbf{s}_t\rangle>>1$ we can approximate the rate expression as
\begin{equation}
        K(\mathbf{s}_t,\mathbf{x}_t) =\underbrace{\langle(\mathbb{I}_N-\mathbf{s}_t),\mathbf{y}_t\rangle}_{\text{Uncoded transmission}} +\underbrace{\left(\frac{\langle \mathbf{s}_t, \mathbb{I}_N\rangle}{M}-1\right)}_{\text{Coded transmission}}
\end{equation}
Note that this approximation makes the rate expression linear in $\mathbf{s}_t$ which makes the computation of $\mathbf{s}_t$ in the step 11 of Algorithm \ref{alg:FTPL} a lot easier. This approximation is good enough for all practical purposes (many a times having the same $\mathbf{s}_t$ as our policy) where $\langle \mathbf{x}_t,\mathbf{s}_t\rangle$ is large. This can be seen from our experiments section.
However one can easily come up with synthetic request patterns where the scheme would achieve a regret that scales linearly with $T$. 

For eg. Consider the scenario where we have $K=4$ users and $M=1$ sized cache at each user and $N=7$ files, namely $\{A,B,C,D,E,F,G\}$ at the server. The requests follows a cyclic structure until $T$. Within a cycle, for one slot the requests are $(A,E.F,G)$, this is followed by $k$ slots where the requests are $(A,B,C,D)$. The linear approximation would then choose a subset that minimizes the approximate cumulative rate $\sum\limits_{i=1}^{t} \left( \langle \mathbf{s}_i, \mathbb{I}_N \rangle - 1 \right) + \langle \mathbb{I}_N - \mathbf{s}_i, \mathbf{y}_i \rangle$. As a result of this we can see the policy eventually will only cache the file $A$ everytime. However the oracle minimizing the exact rate expression from Proposition \ref{proposition:expected_rate} will cache $1/4$ fraction of files $(A,B,C,D)$. Every cycle the total rate incurred the policy using linear approximation for choosing $\mathbf{s}_t$ would be $3(k+1)$ and that incurred by the oracle would be $3k(1-(\frac{3}{4})^4) + 3 +\frac{3}{4}$, The regret incurred per cycle increases with $k$. Since we would have $\mathcal{O}(T)$ such cycles until the horizon, the regret will scale as $\mathcal{O}(T)$.

\label{sec:appstoc}
\subsection{Stochastic request setting: }
In this setting each user $i$ possesses an undisclosed underlying preference distribution across $N$ files, denoted by $p_i = [p_i(1), p_i(2)\cdots p_i(N)]$. This distribution signifies the probability of user $i$ requesting a specific file. This distribution remains the same across time slots and is known to the oracle. As a result of requests being generated from this distribution, $\mathbf{x}_t$ is a random vector $\forall t$. Let $\mathcal{X}$ be the distribution of this random vector over the set of all feasible $\mathbf{x}_t$ $(\left\{\mathbf{x}_t|\langle \mathbf{x}_t,\mathbb{I}_N\rangle = K, \mathbf{x}_t(i)\geq 0\rangle\right\})$ which can be obtained from the distribution of $p_i$, $\forall i$. As a result, the distribution $\mathcal{X}$ is known to the oracle. The expected total rate incurred by a policy $\pi$ is denoted by $\mathcal{K}^{\pi}_S(T)$. The regret incurred by the policy $\pi$ for the stochastic case after $T$ time steps is given by
\begin{equation}
R_\pi^S(T) = \mathcal{K}^{\pi}(T) - T . K_o^S = \sum_{t=1}^{T}\mathbb{E}[K_{\pi}(t) - K_{o}^S] . 
\label{eq:r_stoc}
\end{equation}
Where $K_o^S$ is the expected rate incurred by the static stochastic oracle and is given by
\begin{equation}
    K_o^S = \min_{\mathbf{s}\in \mathcal{S}}K(\mathbf{s})
\end{equation}
where 
\begin{equation*}
     K(\mathbf{s}) = \mathbb{E}_{\mathbf{x}_1\sim \mathcal{X}}\left[K(\mathbf{s},\mathbf{x}_1)\right]
\end{equation*}
Note that $\pi$ here denotes the algorithm that chooses the subset $\mathbf{s}_t$ of files to be cached.
\subsection{Results for Stochastic setting}
Our first result outlines the regret incurred by the policy described in Algorithm \ref{alg:FTPL} w.r.t the stochastic oracle defined in equation (\ref{eq:r_stoc}). The regret incurred in the stochastic case is upper-bounded by a constant, which is consistent with the results obtained in most online learning settings.
\begin{theorem}
    Let $\Delta_{\mathbf{s}} = K(\mathbf{s}) - K_o^S$. The regret incurred by the proposed online policy (Algorithm \ref{alg:FTPL}) and the placement delivery mechanism described above with a learning rate $\eta_t = \alpha\sqrt{t}$  is upper-bounded by 
    \begin{equation}
    R_{\eta}^S(T) \leq \sum_{\mathbf{s}\in \mathcal{S}\setminus \mathbf{s}^*} \frac{64}{\Delta_{\mathbf{s}}}\left((r_{\max}^C)^2+K^2 +\beta\right)
\end{equation}
Here $r_{\max}^C$ is the maximum possible length of a coded transmission over the set of all possible requests $\mathbf{x}_t$ and cache configurations $\mathbf{s}_t$, and $\mathbf{s}^* = \arg\min_{s\in \mathcal{S}}K(s)$ is the cache configuration used by the oracle. $\beta = \alpha^2\max\left\{\frac{M^2}{N},\frac{(N-M)^2}{N} \right\}$
\end{theorem}
\begin{proof}
The proof for this theorem relies on the idea that, given the users have a stationary preference distribution, the policy can learn the distribution of $f(\mathbf{x}_t,\mathbf{s})$ for all $\mathbf{s}$ and $\mathbf{y}_t$ over time after observing an adequate number of request samples. Consequently, it deviates from the Oracle policy with low probability, which is described formally via the lemma \ref{lemma: probub}. 

Let $K_t^{\eta_t}(\mathbf{s}) = \left\langle\left(\mathbf{s}-\frac{M}{N}\mathbb{I}\right), \sum_{i=1}^t f(\mathbf{x}_i,\mathbf{s})-\mathbf{Y}_{t+1} +\eta_t \gamma \right\rangle $. Also, let $\mathcal{B}^t(\mathbf{s})$ be the event that $K_t^{\eta_t}(\mathbf{s})\leq K_t^{\eta_t}(\mathbf{s}^*)$ (an event which will result in a choice of a wrong cache configuration). Note that the event that the policy chooses a cache configuration different from the oracle at time $t$ is a subset of the event $\bigcup_{\mathbf{s}\in\mathcal{S}\setminus \mathbf{s}^*}\mathcal{B}^t(\mathbf{s})$. Then we have
\begin{lemma}
The probability of the event $\mathcal{B}^t(\mathbf{s})$ is upper bounded by
    \begin{equation}
    \mathbb{P}(\mathcal{B}^t(\mathbf{s})) \leq  2\left(e^{-\frac{t\Delta_{\mathbf{s}}^2}{32(r_{\max}^C)^2}}
      +  e^{-\frac{t\Delta_{\mathbf{s}}^2} {32(K)^2}}+
      e^{-\frac{t\Delta_{\mathbf{s}}^2}{32\beta}}\right)
\end{equation}
\label{lemma: probub}
\end{lemma}
\begin{proof}
 We can upper bound the probability of this event as follows
    \begin{align}
        & \mathbb{P}(\mathcal{B}^t(\mathbf{s})) = \mathbb{P}(K_t^{\eta_t}(\mathbf{s})\leq K_t^{\eta_t}(\mathbf{s}^*))  \\
        & = \mathbb{P}\left(\underbrace{\left\langle\left(\mathbf{s}-\frac{M}{N}\mathbb{I}_N\right), \sum_{i=1}^t f(\mathbf{x}_i,\mathbf{s})-\mathbf{Y}_{t+1} +\eta_t \gamma \right\rangle}_{T_a}
        \leq\underbrace{\left\langle\left(\mathbf{s}^*-\frac{M}{N}\mathbb{I}_N\right), \sum_{i=1}^t f(\mathbf{x}_i,\mathbf{s}^*)-Y_{t+1} +\eta_t \gamma \right\rangle}_{T_b}\right) \label {tatb}\\     
        & \leq\mathbb{P}\left(\frac{T_b-tK(\mathbf{s}^*)}{t}\geq \frac{\Delta_{\mathbf{s}}}{2}\right)+ \mathbb{P}\left(\frac{tK(\mathbf{s})-T_a}{t}\geq \frac{\Delta_{\mathbf{s}}}{2}\right)
        \label{eq:split}
    \end{align}
Note that in equation \eqref{tatb}, the expected value of $\mathbb{E}[T_b] = tK(\mathbf{s}^*)$ and the expected value of $\mathbb{E}[T_a] = tK(\mathbf{s})$. Let $T_{aC} = \left\langle\left(\mathbf{s}^*-\frac{M}{N}\mathbb{I}_N\right),\sum_{i=1}^t f(\mathbf{x}_i,\mathbf{s}^*)\right\rangle$, $T_{aU} = \left\langle\left(\mathbf{s}^*-\frac{M}{N}\mathbb{I}_N\right), \mathbf{Y}_{t+1}\right\rangle$ and $T_{aG} = \left\langle\left(\mathbf{s}^*-\frac{M}{N}\mathbb{I}_N\right), \eta_t\gamma\right\rangle$, also $T_{aR} = T_{aC}+T_{aU}$, $T_a = T_{aC}+T_{aU}+T_{aG}$. Similarly, we define $T_{bC}, T_{bU}, T_{bR}$ and $T_{bG}$. Now we have
\begin{align}
     &\mathbb{P}\left(\frac{T_b-tK(\mathbf{s}^*)}{t}\geq \frac{\Delta_{\mathbf{s}}}{2}\right)\\
     & \leq \mathbb{P}\left(\frac{T_{bR}-t\mathbb{E}[T_{bR}]}{t}\geq \frac{\Delta_{\mathbf{s}}}{4}\right)+ \mathbb{P}\left(\frac{T_{bG}-t\mathbb{E}[T_{bG}]}{t}\geq \frac{\Delta_{\mathbf{s}}}{4}\right) \\
     & \leq\mathbb{P}\left(\frac{T_{bC}-t\mathbb{E}[T_{bC}]}{t}\geq \frac{\Delta_{\mathbf{s}}}{8}\right)+ \mathbb{P}\left(\frac{T_{bU}-t\mathbb{E}[T_{bU}]}{t}\geq \frac{\Delta_{\mathbf{s}}}{8}\right) + \mathbb{P}\left(T_{bG}\geq t\frac{\Delta_{\mathbf{s}}}{4}\right)\\
      & \leq e^{-\frac{t\Delta_{\mathbf{s}}^2}{32(r_{\max}^C)^2}}
      +  e^{-\frac{t\Delta_{\mathbf{s}}^2}{32(K)^2}}+\mathbb{P}\left(T_{bG}\geq t\frac{\Delta_{\mathbf{s}}}{4}\right)\label{eq:hoeff}\\
      & \leq  e^{-\frac{t\Delta_{\mathbf{s}}^2}{32(r_{\max}^C)^2}}
      +  e^{-\frac{t\Delta_{\mathbf{s}}^2} {32(K)^2}}+
      e^{-\frac{t\Delta_{\mathbf{s}}^2}{32\beta}}
      \label{eq:gausbound}
\end{align}
Equation \eqref{eq:hoeff} comes from Hoeffding inequality and the fact $T_{bC} \in [0, r_{\max}^C]$ (Its the coded rate incurred when the cache configuration is $\mathbf{s}^*$) and $T_{bU} \in \left[-\frac{M}{N}K, \left(1-\frac{M}{N}\right)K\right]$. Equation \eqref{eq:gausbound} comes from the upper deviation inequality for a Gaussian random variable $X\sim \mathcal{N}(\mu,\sigma^2)$ which says $\mathbb{P}[\mathbf{X}\geq \mu+t]\leq \exp(-t^2/2\sigma^2)$ and the max variance of $T_{bG}$ is $\max\left\{\frac{M^2}{N},\frac{(N-M)^2}{N} \right\}\eta_t^2 = \beta t$. ($\eta_t = \alpha \sqrt{t}$), . Similarly, one can upperbound the second term in equation \eqref{eq:split} to get the desired result.
\end{proof}
 Let $\Delta_{\mathbf{s}} = K(\mathbf{s}) - K_o^S$.
 Let $K_t^{\eta_t}(\mathbf{s}) = \left\langle\left(s-\frac{M}{N}\mathbb{I}_N\right), \sum_{i=1}^t f(\mathbf{x}_i,\mathbf{s})-\mathbf{Y}_{t+1} +\eta_t \gamma \right\rangle $. We have
 \begin{align}
      R_{\eta}^S(T)&= \sum_{t=1}^T\mathbb{E}[K_{\pi}(t)-K_o^S(t)]  = \sum_{t=1}^T\sum_{\mathbf{s}\in\mathcal{S}\setminus \mathbf{s}^*}\mathbb{E}[(K_{\mathbf{s}}(t)-K_o^S(t))\mathbb{I}(\mathbf{s}_t = s)] \label{eq:ind}\\
      & = \sum_{t=1}^T\sum_{s\in\mathcal{S}\setminus s^*} \Delta_{\mathbf{s}_t} \mathbb{P}(\mathbf{s}_t= \mathbf{s})\\
     & \leq \sum_{t=1}^T \sum_{\mathbf{s}\in \mathcal{S}\setminus \mathbf{s}^*} \Delta_{\mathbf{s}} \mathbb{P}(K_t^{\eta_t}(\mathbf{s})\leq K_t^{\eta_t}(\mathbf{s}^*)) =  \sum_{t=1}^T \sum_{\mathbf{s}\in \mathcal{S}\setminus \mathbf{s}^*} \Delta_{\mathbf{s}} \mathbb{P}(\mathcal{B}^t(\mathbf{s}))
 \end{align}
In equation \eqref{eq:ind} $K_{\mathbf{s}}(t) = \left\langle\left(\mathbf{s}-\frac{M}{N}\mathbb{I}_N\right), f(\mathbf{x}_t,\mathbf{s})-\mathbf{y}_t\right\rangle +  h(\mathbf{x}_t) $ and $K_o(t) = \left\langle\left(\mathbf{s}^*-\frac{M}{N}\mathbb{I}_N\right), f(\mathbf{x}_t,\mathbf{s}^*)-\mathbf{y}_t\right\rangle +  h(\mathbf{x}_t)$. Note that the random variables $K_{\mathbf{s}}(t)-K_o^S(t)$ (expected value $\Delta_{\mathbf{s}}$) and $\mathbb{I}(\mathbf{s}_t=\mathbf{s})$ (expected value $\mathbb{P}(\mathbf{s}_t=\mathbf{s})$) are independent. Also, $\mathcal{B}^t(\mathbf{s})$ is the event that $K_t^{\eta_t}(\mathbf{s})\leq K_t^{\eta_t}(\mathbf{s}^*)$ (an event which results in a choice of a wrong cache configuration). From the lemma below, we have
\begin{align}
    & \sum_{t=1}^T \sum_{\mathbf{s}\in \mathcal{S}\setminus \mathbf{s}^*} \Delta_{\mathbf{s}} \mathbb{P}(\mathcal{B}^t(\mathbf{s}))\\
    & \leq  \sum_{\mathbf{s}\in \mathcal{S}\setminus\mathbf{s}^*}  \sum_{t=1}^T 2\Delta_\mathbf{s} \left(e^{-\frac{t\Delta_{\mathbf{s}}^2}{32(r_{\max}^C)^2}}
      +  e^{-\frac{t\Delta_{\mathbf{s}}^2} {32(K)^2}}+
      e^{-\frac{t\Delta_{\mathbf{s}}^2}{32\beta}}\right)\\
      &\leq \sum_{\mathbf{s}\in \mathcal{S}\setminus \mathbf{s}^*}  \sum_{t=1}^{\infty} 2\Delta_{\mathbf{s}} \left(e^{-\frac{t\Delta_{\mathbf{s}}^2}{32(r_{\max}^C)^2}}
      +  e^{-\frac{t\Delta_{\mathbf{s}}^2} {32(K)^2}}+
      e^{-\frac{t\Delta_{\mathbf{s}}^2}{32\beta}}\right)\\
      & \leq \sum_{\mathbf{s}\in \mathcal{S}\setminus \mathbf{s}^*} 2\Delta_\mathbf{s} \left(\frac{\exp(-\frac{\Delta_{\mathbf{s}}^2}{32(r_{\max}^C)^2})} {1-\exp(-\frac{\Delta_{\mathbf{s}}^2}{32(r_{\max}^C)^2})} + \frac{\exp(-\frac{\Delta_{\mathbf{s}}^2} {32(K)^2})} {1-\exp(-\frac{\Delta_{\mathbf{s}}^2} {32(K)^2})}+ \frac{\exp(-\frac{\Delta_{\mathbf{s}}^2}{32\beta})} {1-\exp(-\frac{\Delta_{\mathbf{s}}^2}{32\beta})}\right)\\
      & \leq \sum_{\mathbf{s}\in \mathcal{S}\setminus \mathbf{s}^*} \frac{64}{\Delta_{\mathbf{s}}}\left((r_{\max}^C)^2+K^2 +\beta\right)
\end{align}
Last step follows since $e^{-x}/(1-e^{-x})\leq \frac{1}{x}$
\end{proof}

\subsection{Switching Cost: Stochastic requests}
\label{app: stocswitch}
\begin{theorem}
    The number of switches up to time $T$ for the Algorithm \ref{alg:FTPL} for the stochastic setting defined in equation \eqref{eq:r_stoc} can be upper bounded as
    \begin{equation}
        \sum_{t=1}^{T-1}\mathbb{E}[\mathbb{I}(\mathbf{s}_{t+1}\neq \mathbf{s}_{t}] \leq \sum_{s\in \mathcal{S}\setminus \mathbf{s}^*} \frac{64}{\Delta_{\mathbf{s}}^2}\left((r_{\max}^C)^2+K^2 +\beta\right)
    \end{equation}
\end{theorem}
\begin{proof}
The idea of bounding the number of switches in cache configuration until time $t$ here is again based on the idea that eventually, with high probability, the policy will choose the same cache configurations as the oracle. Let the sequence of cache configurations cache upto time $t$ be $\mathbf{\mathbf{s}}_{\text{Seq}} = (\mathbf{\mathbf{s}}_1, \mathbf{\mathbf{s}}_2\cdots \mathbf{\mathbf{s}}_T)$.  Now consider the sequence  $\mathbf{\mathbf{s}}_{\text{Seq}}^M = (\mathbf{\mathbf{s}}_1, \mathbf{s}^*, \mathbf{\mathbf{s}}_2, \mathbf{s}^*,\cdots \mathbf{\mathbf{s}}_T, \mathbf{s}^*)$. Now observe that the number of switches in the sequence $\mathbf{s}_{\text{Seq}}$ will be less than the number of switches in the sequence $\mathbf{s}_{\text{Seq}}^M$. One can further observe that the number of switches is the sequence $\mathbf{s}_{\text{Seq}}^M$ is less than $2\sum_{t=1}^T \mathbb{I}(\mathbf{s}_t\neq \mathbf{s}^*)$. Thus we have
\begin{align}
    \sum_{t=1}^{T-1}\mathbb{E}[\mathbb{I}(\mathbf{s}_{t+1}\neq \mathbf{s}_{t}] \leq& \sum_{t=1}^{T-1}\left(\mathbb{E}[\mathbb{I}(\mathbf{s}_t\neq \mathbf{s}^*)]+\mathbb{E}[\mathbb{I}(\mathbf{s}^*\neq\mathbf{s}_{t+1})]\right)\\
    \leq & 2\sum_{t=1}^{T-1}\mathbb{E}[\mathbb{I}(\mathbf{s}_t\neq \mathbf{s}^*)]\\
    = & 2\sum_{t=1}^{T}\mathbb{P}(\mathbf{s}_t\neq \mathbf{s}^*)\\
    \leq & 2\sum_{t=1}^{T}\mathbb{P}\left(\bigcup_{\mathbf{s}\in\mathcal{S}\setminus \mathbf{s}^*}\mathcal{B}^t(\mathbf{s})\right)\label{eq: badub}\\ 
    \leq & 2\sum_{t=1}^{T}\sum_{\mathbf{s}\in\mathcal{S}\setminus \mathbf{s}^*}\mathbb{P}\left(\mathcal{B}^t(\mathbf{s})\right)\\
    \leq & 2\sum_{\mathbf{s}\in \mathcal{S}\setminus \mathbf{s}^*}  \sum_{t=1}^{\infty}  \left(e^{-\frac{t\Delta_s^2}{32(r_{\max}^C)^2}}
      +  e^{-\frac{t\Delta_{\mathbf{s}}^2} {32(K)^2}}+
      e^{-\frac{t\Delta_{\mathbf{s}}^2}{32\beta}}\right)\label{eq:lemmmaimpl}\\
     \leq & 2 \sum_{\mathbf{s}\in \mathcal{S}\setminus \mathbf{s}^*} \left(\frac{\exp(-\frac{\Delta_{\mathbf{s}}^2}{32(r_{\max}^C)^2})} {1-\exp(-\frac{\Delta_{\mathbf{s}}^2}{32(r_{\max}^C)^2})} + \frac{\exp(-\frac{\Delta_{\mathbf{s}}^2} {32(K)^2})} {1-\exp(-\frac{\Delta_{\mathbf{s}}^2} {32(K)^2})}+ \frac{\exp(-\frac{\Delta_{\mathbf{s}}^2}{32\beta})} {1-\exp(-\frac{\Delta_{\mathbf{s}}^2}{32\beta})}\right)\\
       \leq & \sum_{\mathbf{s}\in \mathcal{S}\setminus \mathbf{s}^*} \frac{64}{\Delta_{\mathbf{s}}^2}\left((r_{\max}^C)^2+K^2 +\beta\right)
\end{align}
Equation \eqref{eq: badub} comes from the fact that the event under which the policy chooses a cache configuration different from the oracle at time $t$ as ($\{\mathbf{s}_t\neq \mathbf{s}^*\}$) is a subset of the event $\bigcup_{\mathbf{s}\in\mathcal{S}\setminus \mathbf{s}^*}\mathcal{B}^t(\mathbf{s})$. Equation \eqref{eq:lemmmaimpl} comes from the lemma \ref{lemma: probub}. Last step follows since $e^{-x}/(1-e^{-x})\leq \frac{1}{x}$
\end{proof}
\subsection{Restricted Switching: Stochastic requests}
\label{app: resstoc}
Switching the cache configuration is only allowed in slots $t_i$ for $i\in\{1,2\cdots L\}$ where $t_k = \sum_{i=1}^{k}l_i$ where $l_i$ are the inter-switch periods. The policy (Algorithm \ref{alg:FTPL}) is used here whenever we have a switching slot. i.e., The cache placement step is done only during the switching slot, and the delivery step is performed in every slot. By convention, we allow switching after the last slot i.e., $\sum_{k=1}^L l_k = t_{L}= T$
\begin{theorem}
Let $R^S_{(\eta, \text{R})}(T)$ be the regret incurred by the policy (algorithm \ref{alg:FTPL}) in the stochastic restricted switching case with a horizon $T$. Then, the regret incurred by this policy in the restricted switching scenario with a learning rate $\alpha\sqrt{t}$ is upper-bounded as
\begin{equation}
    R^S_{(\eta, \text{R})}(T) \leq r_{\max}l_1 + \sum_{k=1}^{L}\sum_{\mathbf{s}\in \mathcal{S}\setminus \mathbf{s}^*} 2 l_k \Delta_{\mathbf{s}} \left(e^{-\frac{t_k\Delta_{\mathbf{s}}^2}{32(r_{\max}^C)^2}}
      +  e^{-\frac{t_k\Delta_{\mathbf{s}}^2} {32(K)^2}}+
      e^{-\frac{l_k\Delta_{\mathbf{s}}^2}{32\beta}}\right)
\end{equation}
\end{theorem}

\begin{proof}
 If the algorithm chooses a cache configuration different from the oracle at time $t_k$ then it incurs non-zero regret w.r.t. the oracle from time $t_k$ to time $t_{k+1}-1$. Let $R_k^S$ be the regret incurred by the oracle between $t_k$ and $t_{k+1}-1$. Let $\mathbf{s}\neq \mathbf{s}^*$ be the cache configuration chosen at time $t_i$. Then we have
 \begin{align}
     R_K^S  &= \sum_{t=t_k}^{t_{k+1}-1}\mathbb{E}[K_{\pi}(t)-K_o(t)]  \\
     &= \sum_{t=t_k}^{t_{k+1}-1}\sum_{\mathbf{s}\in\mathcal{S}\setminus \mathbf{s}^*}\mathbb{E}[(K_{\mathbf{s}}(t)-K_o(t))\mathbb{I}(\mathbf{s}_t = \mathbf{s})] \label{eq:indl}\\
     &= l_k\sum_{\mathbf{s}\in\mathcal{S}\setminus \mathbf{s}^*}\mathbb{E}[(K_{\mathbf{s}_{t_k}}-K_o(t_k))\mathbb{I}(\mathbf{s}_{t_k} = \mathbf{s})] \label{eq:samecc}\\     
     & = l_k\sum_{\mathbf{s}\in\mathcal{S}\setminus \mathbf{s}^*} \Delta_{\mathbf{s}_{t_k}} \mathbb{P}(\mathbf{s}_{t_k}= \mathbf{s})\label{eq:samecc2}\\
     & \leq l_k \sum_{\mathbf{s}\in \mathcal{S}\setminus \mathbf{s}^*} \Delta_{\mathbf{s}} \mathbb{P}\left(K_{t_k}^{\eta_{t_k}}(s)\leq K_{t_k}^{\eta_{t_k}}(\mathbf{s}^*)\right) =  l_k \sum_{\mathbf{s}\in \mathcal{S}\setminus \mathbf{s}^*} \Delta_{\mathbf{s}} \mathbb{P}(\mathcal{B}^{t_k}(\mathbf{s}))\\
     & \leq  \sum_{\mathbf{s}\in \mathcal{S}\setminus \mathbf{s}^*} 2 l_k \Delta_{\mathbf{s}} \left(e^{-\frac{t_k\Delta_{\mathbf{s}}^2}{32(r_{\max}^C)^2}}
      +  e^{-\frac{t_k\Delta_{\mathbf{s}}^2} {32(K)^2}}+
      e^{-\frac{t_k\Delta_{\mathbf{s}}^2}{32\beta}}\right) \label{eq:itsover}
 \end{align}
 Equation \eqref{eq:samecc} the cache configuration remains fixed for these slots. Equation \eqref{eq:samecc2} comes from the independence of the random variables within the expectation.
 The equation \eqref{eq:itsover} comes from lemma \ref{lemma: probub}. Upper bounding the regret before the first switching slot by $l_1r_{\max}$, we have the required result.
\end{proof}

\end{document}